\documentclass[preprint,showpacs,showkeys,preprintnumbers,amsmath,amssymb,eqsecnum]{revtex4}

\usepackage{amsmath,amssymb,amscd,amsbsy,amsgen,amsopn,amstext,amsxtra}

\usepackage{graphicx}
\usepackage{dcolumn}
\usepackage{bm}
\usepackage[mathscr]{eucal}

\begin{document}

\renewcommand{\thefootnote}{*\arabic{footnote}}

\preprint{ }

\setlength{\baselineskip}{6.57mm}

\title{A Supersphere formulation of Yang-Mills theory on sphere}

\author{\large Rabin Banerjee}
\email{rabin@bose.res.in}

\affiliation{S.N. Bose National Centre for Basic Sciences, 
JD Block, Sector III, Salt Lake City, Kolkata 700 098, India \\ 
{\rm and} 
Instiut f\"{u}r Theoretische Physik, Universit\"{a}t zu K\"{o}ln, \\ 
Z\"{u}lpicher Strasse 77, 50937 Cologne, Germany.}

\author{\large Shinichi Deguchi}
\email{deguchi@phys.cst.nihon-u.ac.jp}

\affiliation{Institute of Quantum Science, College of Science and Technology, 
Nihon University, Chiyoda-ku, Tokyo 101-8308, Japan} 

\author{~}

\begin{abstract}
A superfield approach to the Becchi-Rouet-Stora-Tyutin (BRST) formalism 
for the Yang-Mills theory on an $n$-dimensional unit sphere, $S_1^{n}$,  
is developed in a manifestly covariant manner based on 
the rotational supersymmetry characterized by the supergroup ${\rm OSp}(n+1|2)$. 
This is done by employing an $(n+2)$-dimensional unit supersphere, $S_1^{n|2}$, 
parametrized by $n$ commutative and 2 anticommutative 
coordinate variables 
so that it includes $S_1^{n}$ as a subspace and realizes the ${\rm OSp}(n+1|2)$  
supersymmetry.  
In this superfield formulation, referred to as the {\em supersphere formulation},  
the so-called horizontality condition is concisely expressed  
in terms of the rank-3 field strength tensor of a Yang-Mills superfield on  
$S_1^{n|2}$. The supersphere formulation completely covers 
the BRST gauge-fixing procedure for the Yang-Mills theory on $S_1^{n}$ provided by us 
[R. Banerjee and S. Deguchi, Phys. Lett. B 632 (2006) 579, arXiv:hep-th/0509161]. 
Furthermore, this formulation admits the (massive) Curci-Ferrari model defined on $S_1^{n}$, 
describing the gauge-fixing and mass terms on $S_1^{n}$ together  
as a mass term on $S_1^{n|2}$. 
\end{abstract}

\pacs{11.15.-q, 11.10.Kk, 11.30.Pb}

\keywords{supersphere, superfield, horizontality condition, 
BRST symmetry, Curci-Ferrari model} 

\maketitle

\newpage


\section{\label{sec:level1}Introduction}

Manifestly O($n+1$) covariant formulation of gauge theories on an $n$-dimensional 
sphere, $S^{n}$, has been studied in various contexts 
\cite{Adl,DS,Sho,JR,Ore,NS,Ban,BD} 
since Adler developed the O(5) covariant formulation of 
QED (quantum electrodynamics) on $S^{4}$ \cite{Adl}. 
In earlier studies of QED and the Yang-Mills theory 
formulated in manifestly O($n+1$) covariant forms \cite{DS,Sho},   
an unconventional gauge-fixing term was introduced into the actions 
in such a way that it leads to 
the gauge-fixing condition proposed by Adler \cite{Adl}. 
The associated Faddeev-Popov (FP) ghost term was also found in somewhat 
complicated manner. 
However, the Becchi-Rouet-Stora-Tyutin (BRST) symmetry and 
the BRST invariance principle were not considered there.

Recently, the gauge-fixing procedure based on the 
BRST invariance principle  
(or simply BRST gauge-fixing procedure) \cite{KU, NO} 
has been applied to 
the Yang-Mills theory on $S^n$ 
in a manner such that manifestly O($n+1$) covariance is maintained \cite{BD}. 
In this approach, 
the gauge-fixing condition proposed by Adler was generalized 
to incorporate a gauge parameter. 
However, the generalized Adler condition was not used in its own form,  
because this condition has an extra free index and 
hence is not appropriate for the ordinary first-order formalism 
of gauge fixing \cite{NL}. To avoid this difficulty, 
the BRST gauge-fixing procedure for the Yang-Mills theory on $S^n$  
adopted a gauge-fixing condition that is equivalent to 
the generalized Adler condition, but does not have extra free indices. 
The equivalence of the two conditions was proven in an elegant manner \cite{BD},  
and consequently the condition adopted was recognized to be an alternative form 
of the generalized Adler condition. 
With the appropriate gauge-fixing condition, 
the sum of gauge-fixing and FP ghost terms was defined as   
a coboundary term with respect to the BRST transformation 
satisfying the nilpotency property. 
Then, it was shown that the total action with these gauge-fixing and FP ghost terms 
yields the field equations on $S^n$ that have manifestly O($n+1$) covariant 
or invariant forms. 
Also, it was demonstrated, with the aid of conformal Killing vectors \cite{Ban},  
that the field equations on $S^n$ reduce, in the large radius limit of $S^n$,  
to corresponding equations in the Yang-Mills theory on $n$-dimensional  
Euclidian space.

Having established the BRST formalism for the Yang-Mills theory on $S^n$,  
it is natural to ask how this formalism is described in geometrical terms 
of superspace. 
For the Yang-Mills theory on the flat space, there have been several  
sorts of superfield approaches to the BRST formalism \cite{sff,Fuj1,JMD}. 
The approach developed in Refs. \cite{sff} 
begins with the flat superspace with two anticommutative   
coordinate variables and introduces a generalized Yang-Mills field, 
referred to as a Yang-Mills superfield, 
into the superspace. The (anti-)BRST transformation rules of 
the ordinary Yang-Mills and FP (anti-)ghost fields are realized in this approach as 
the so-called horizontality condition imposed on the field strength 
of the Yang-Mills superfield. 
(Superfield approaches without the 
horizontality condition have been developed in Refs. \cite{Fuj1,JMD}.) 
In the superfield formulation in Refs. \cite{sff}, 
the nilpotency and anticommutativity properties 
of the BRST and anti-BRST transformations are understood from 
the anticommuting property of the anticommutative coordinate variables. 
Also, the gauge-fixing term that has the form of a double coboundary term 
with respect to both the BRST and anti-BRST transformations 
can be expressed as a mass term for the Yang-Mills superfield.

The purpose of the present paper is to develop  
a superfield approach to the BRST formalism for the Yang-Mills theory on $S^n$. 
To this end, we first note the fact that  
the Yang-Mills theory on $S^n$ treats angular momentum operators 
as more fundamental operators than usual derivative operators, 
because translations on $S^n$ are performed by rotations. 
Correspondingly, the field strength of the Yang-Mills field on $S^n$ is 
defined as a totally antisymmetric tensor of rank 3, rather than the usual 
field strength tensor of rank 2 \cite{Sho,JR,Ore,NS,Ban,BD}. 
Therefore it follows that the superspace generalization of  
the Yang-Mills theory on $S^n$ necessarily involves   
the rank-3 field strength tensor for the Yang-Mills superfield on a superspace. 
In the ordinary superfield formulation mentioned above, 
the horizontality condition is imposed on the field strength 
of the Yang-Mills superfield. 
Following this, in the present approach, 
we impose a horizontality condition on the rank-3 field strength tensor 
of the Yang-Mills superfield.

Now it is clear that the flat superspace is not appropriate for 
the superfield formulation of the Yang-Mills theory on $S^n$. 
A desirable superspace (or supermanifold)    
must include $S^n$ as a subspace, and 
furthermore it must possess supersymmetry that is a generalization 
of the rotational symmetry characterized by the orthogonal group ${\rm O}(n+1)$. 
Such a superspace has already been considered in 
some different contexts \cite{ND, susp}, and nowadays it is known as 
the {\em supersphere}.   
The present paper employs the $(n+2)$-dimensional supersphere,  
$S^{n|2}$, which is parametrized by $n$ commutative and 2 anticommutative 
coordinate variables. 
As expected, $S^{n|2}$ includes $S^n$, and 
possesses the rotational supersymmetry characterized by 
the orthosymplectic supergroup ${\rm OSp}(n+1|2)$ \cite{FK,DeW}. 
Generalized angular momentum operators are realized on $S^{n|2}$ 
as generators of ${\rm OSp}(n+1|2)$, with which 
we can define the rank-3 field strength tensor for the Yang-Mills superfield on $S^{n|2}$.

In our superfield formulation based on the supersphere $S^{n|2}$  
(or simply {\em supersphere formulation}),  
the horizontality condition is thus imposed on 
the rank-3 field strength tensor of the Yang-Mills superfield on $S^{n|2}$. 
As will be seen later, the horizontality condition takes a concise form,   
$\hat{\mathcal{F}}_{a\beta\gamma}=0$. 
(Here, $a$ is an index for the commutative coordinates, while 
$\beta$ and $\gamma$ are indices for the anticommutative coordinates.) 
This yields relations among some of the component fields on $S^{n}$  
that are given as expansion coefficients of the Taylor series expansion of the Yang-Mills 
superfield with respect to the anticommutative coordinate variables. 
The zeroth-order terms of this Taylor series are eventually identified with  
the Yang-Mills and FP (anti-)ghost fields on $S^{n}$, up to constants. 
Their (anti-)BRST transformation rules can be derived from the 
relations among the component fields.  
The BRST transformation rules obtained through this procedure are identical to 
those found in a previous paper \cite{BD}.

The action for the Yang-Mills field on 
$S^n$ is defined from the Yang-Mills field strength tensor of rank-3 
\cite{Sho,JR,Ore,NS,Ban,BD}.   
This action can be expressed as an action for the Yang-Mills superfield 
on $S^{n|2}$ that is written in terms of 
its rank-3 field strength tensor supplemented with the horizontality condition.  
The gauge-fixing term on $S^n$ that takes the form of a double coboundary term 
with respect to the BRST and anti-BRST transformations can be expressed as 
a {\em generalized} mass term for the Yang-Mills superfield on $S^{n|2}$, 
not as the (naive) mass term for it. 
With a suitable choice of constant parameters, the generalized mass term 
can also reduce to the sum of the double-coboundary gauge-fixing term 
and a mass term for the Yang-Mills and (anti-)FP ghost fields on $S^n$.  
The mass term found here is shown to be the Curci-Ferrari mass term \cite{CF1, CF2} 
defined on $S^n$.  In this sense, 
the supersphere formulation admits the (massive) Curci-Ferrari model on $S^n$.  
In a particular case, the generalized mass term becomes the naive mass term 
for the Yang-Mills superfield on $S^{n|2}$. Remarkably,  
this term yields the Curci-Ferrari mass term with definite 
mass values that depend only on space dimension $n$. 
This can be understood as a reflection of the ${\rm OSp}(n+1|2)$ 
symmetry of the naive mass term.

The present paper is organized as follows: 
Section 2 provides a brief review of the manifestly ${\rm O}(n+1)$ covariant formulation 
of the Yang-Mills theory on an $n$-dimensional {\em unit} sphere, $S_{1}^{n}$  
{\footnotemark[1]}. 
%
\footnotetext[1]{In the present paper, the radii of $S^{n}$ and $S^{n|2}$ 
are assumed to be {\em unity} for simplicity.   
This choice does not lose generalities.} 
%
The BRST gauge-fixing procedure for this theory is explained in detail. 
In section 3, an $(n+2)$-dimensional {\em unit} supersphere, $S_{1}^{n|2}$,  
is defined based on Refs. \cite{ND}; 
also, embedding $S_{1}^{n}$ in $S_{1}^{n|2}$ is carried out so that $S_{1}^{n}$ can be 
a subspace of $S_{1}^{n|2}$. 
Section 4 introduces a Yang-Mills superfield into $S_{1}^{n|2}$ 
and treats its component fields defined on $S_{1}^{n}$. 
The tensor components of the rank-3 field strength tensor of this Yang-Mills superfield 
are expressed in terms of the component fields. 
Section 5 analyses the above-mentioned horizontality condition  
$\hat{\mathcal{F}}_{a\beta\gamma}=0$, showing  
that it indeed yields the (anti-)BRST transformation rules of the 
relevant fields on $S_{1}^{n}$. 
Section 6 presents a modified Yang-Mills action on $S_{1}^{n|2}$ that 
turns out to be the Yang-Mills action on $S_{1}^{n}$. 
Section 7 considers two gauge-fixing terms expressed as mass terms for 
the Yang-Mills superfield on $S_{1}^{n|2}$. 
It is demonstrated there that one of the gauge-fixing terms, 
with a vanishing condition of constant parameters, 
turns out to be a generalization 
of the gauge-fixing term proposed in  Ref. \cite{BD}. 
Section 8 shows that the supersphere formulation admits 
the Curci-Ferrari model on $S_{1}^n$. 
Section 9 is devoted to a summary and discussion.


\section{\label{sec:level1}Yang-Mills theory on sphere (A brief review)}

In this section, we briefly review a manifestly O($n+1$) covariant formulation of 
the Yang-Mills theory on an $n$-dimensional sphere \cite{Adl,JR,Ore,NS,Ban,BD} 
for the convenience of later studies.

Let us consider an $n$-dimensional unit sphere $S_{1}^n$ 
embedded in $(n+1)$-dimensional Euclidean space ${\bf R}^{n+1}$. 
The sphere $S_{1}^n$ is characterized by the following constraint 
imposed on Cartesian coordinates $(r_a)$ $(a=1,2,\ldots, n+1)$ on ${\bf R}^{n+1}\,$: 
\begin{align}
r_a r_a=r_{\mu} r_{\mu} +(r_{n+1})^2 =1 \,.
\label{2.0}
\end{align}
We can use $(r_\mu)$  $(\mu=1,2,\ldots, n\,;\,0\leq r_\mu r_\mu \leq 1)$ as local 
coordinates on $S_{1}^n$, 
treating $r_{n+1}=\pm\sqrt{1-r_{\mu} r_{\mu}}$ as a dependent variable \footnotemark[2]. 
%
\footnotetext[2]{The indices $a, b, c, d,$ and $e$ run from $1$ to $n+1$, 
while the indices $\mu$ and $\nu$ run from $1$ to $n$.} 
%
In terms of the independent variables $(r_\mu)$, the generators of O($n+1$) 
(or the angular momentum operators) $L_{ab}$ read 
\begin{align}
&L_{\mu\nu}=-i( r_\mu \partial_\nu -r_\nu \partial_\mu) \,, \quad 
\partial_\mu \equiv \frac{\partial}{\partial r_\mu} \,,
\label{2.1}
\\ 
&L_{\mu (n+1)}=-L_{(n+1)\mu}=ir_{n+1}\partial_{\mu} \,, 
\label{2.2}
\end{align}
or more concisely 
\begin{align}
L_{ab}=-i( r_a \partial_b -r_b \partial_a )\,, \quad 
\partial_a\equiv \delta_{a\mu}\partial_{\mu} \,.
\label{2.3}
\end{align}
Noting that 
\begin{align}
\frac{\partial r_{n+1}}{\partial r_{\mu}}=-\frac{r_{\mu}}{r_{n+1}} \,, 
\label{2.4}
\end{align}
we can show that the generators in Eqs. (\ref{2.1}) and (\ref{2.2}) 
satisfy the commutation relations of the O($n+1$) Lie algebra,    
\begin{align}
[L_{ab}, L_{cd}]
=i ( \delta_{ac}L_{bd} -\delta_{bc}L_{ad} -\delta_{ad}L_{bc} +\delta_{bd}L_{ac}) \,.
\label{2.5}
\end{align}

Let $\hat{A}_{a}$ be a (Hermitian) Yang-Mills field on $S^{n}_{1}$ that takes values in a 
compact semisimple Lie algebra ${\frak g}$; $\hat{A}_{a}$ can be expanded as  
$\hat{A}_{a}=\sum_{i=1}^{{\rm dim}{\frak g}}\hat{A}_{a}^{i}T_{i}$ in 
terms of the Hermitian basis $\{ T_{i} \}$ of ${\frak g}$ which satisfy the commutation relations 
$[T_{i}, T_{j}]=if_{ij}{}^{k}T_{k} $ and 
the normalization conditions $\mathrm{Tr}(T_i T_j)=\delta_{ij}$ \footnotemark[3].  
%
\footnotetext[3]{The indices $i, j,$ and $k$ run from $1$ to ${\rm dim}{\frak g}$.}  
%
We can regard $\hat{A}_{a}$ as a function of the independent variables $(r_{\mu})$. 
The Yang-Mills field $\hat{A}_{a}$ is assumed to live on the tangent space, 
$\mathfrak{T}_P S^{n}_{1}$, at a point 
$P(r_{\mu})$ on $S_1^n$ by imposing the transversality condition 
\begin{align}
r_a \hat{A}_{a}=0 \,.
\label{2.6}
\end{align}
This implies that one component of $(\hat{A}_{a})$, for instance $\hat{A}_{n+1}$, depends 
on the other components, such as $\hat{A}_{n+1}=-(r_\mu \hat{A}_{\mu})/r_{n+1}$. 
The infinitesimal gauge transformation of $\hat{A}_{a}$ is given by \cite{Ban,BD}
\begin{align}
\delta_{\lambda} \hat{A}_a 
&= ir_b \mathcal{L}_{ba} \lambda =P_{ab}\hat{D}_{b} \lambda \,, 
\label{2.7}
\end{align}
where $\lambda$ is an infinitesimal function taking values in ${\frak g}$, 
$\mathcal{L}_{ab}$ are covariantized angular momentum operators  
\begin{align}
\mathcal{L}_{ab}&\equiv L_{ab}-[r_a \hat{A}_b -r_b \hat{A}_a, \;\;\, ]
=-i(r_a \hat{D}_b -r_b \hat{D}_a) \,,
\label{2.8}
\end{align}
while $P_{ab}$ and $\hat{D}_{a}$ are the tangential projection operator and the covariant 
derivative, respectively: 
\begin{align}
P_{ab} &\equiv \delta_{ab}-r_a r_b \,, 
\label{2.9}
\\ 
\hat{D}_a &\equiv \partial_a -i[ \hat{A}_a, \;\;\, ] \,.
\label{2.10}
\end{align}
The projection operator $P_{ab}$ in Eq. (\ref{2.7}) guarantees that the Yang-Mills field 
transformed according to the rule (\ref{2.7}), i.e.,  
$\hat{A}_a +\delta_{\lambda} \hat{A}_a$, lives on the tangent space ${\frak T}_P S^{n}_{1}$.

As has been emphasized in the literature \cite{JR,Ore,Ban,BD}, 
the field strength of $\hat{A}_{a}$ can be written in a manifestly 
O($n+1$) covariant form: 
\begin{subequations}
\label{2.11}
\begin{align}
\hat{F}_{abc}&= i( L_{ab}\hat{A}_{c}-r_a [\hat{A}_{b}, \hat{A}_{c} ] \big)
\nonumber \\ 
&\,\quad +\mbox{cyclic permutations in $(a, b, c)$}
\label{2.11a} 
\\ 
&= r_a \hat{F}_{bc} +r_b \hat{F}_{ca} +r_c \hat{F}_{ab} \,, 
\label{2.11b}
\end{align}
\end{subequations}
where $\hat{F}_{ab}$ is defined by 
\begin{align}
\hat{F}_{ab}=\partial_a \hat{A}_b -\partial_b \hat{A}_a
-i [\hat{A}_{a}, \hat{A}_{b} ] \,. 
\label{2.12}
\end{align}
Althougth $\hat{F}_{ab}$ transforms inhomogeneously under 
the gauge transformation (\ref{2.7}), i.e., 
\begin{align}
\delta_{\lambda} \hat{F}_{ab}&= -i[ \hat{F}_{ab}, \lambda] 
+r_a \bigg( \hat{D}_{b}+\frac{1}{r_{n+1}}\delta_{b (n+1)}
\bigg)(r_{\mu}\partial_{\mu} \lambda) 
\nonumber \\ 
&\quad \,-r_b \bigg( \hat{D}_{a}+\frac{1}{r_{n+1}}\delta_{a (n+1)}
\bigg)(r_{\mu}\partial_{\mu} \lambda) \,, 
\label{2.13} 
\end{align}
the rank-3 tensor $\hat{F}_{abc}$ 
transforms homogeneously \cite{BD}:  
\begin{align}
\delta_{\lambda} \hat{F}_{abc}= -i[ \hat{F}_{abc}, \lambda] \,.
\label{2.14}
\end{align}
Thus $\hat{F}_{abc}$ has the property of field strength. 
With the field strength $\hat{F}_{abc}$, the Yang-Mills action for $\hat{A}_{a}$ 
is written as 
\begin{align}
S_{\rm YM}=\int d^{n} \varOmega \bigg[ -{1\over12} \mathrm{Tr}(\hat{F}_{abc} \hat{F}_{abc}) 
\bigg] \,, 
\label{2.15}
\end{align}
where $d^{n}\varOmega$ is an invariant measure on $S_1^n$ defined by  
\begin{align}
d^{n} \varOmega\equiv \frac{1}{|r_{n+1}|} \prod_{\mu=1}^{n} dr_{\mu} \,. 
\label{2.16}
\end{align} 
Obviously, the action $S_{\rm YM}$ is gauge invariant.
The variation of $S_{\rm YM}$ with respect to $\hat{A}_{a}$ gives a Yang-Mills equation 
of the form $\mathcal{L}_{ab} \hat{F}_{abc}=0$.

In order to investigate quantum-theoretical properties of the Yang-Mills theory on 
$S_1^n$, it is necessary to introduce a suitable gauge-fixing condition to the theory. 
The Adler condition, $iL_{ab} \hat{A}_b =\hat{A}_a$, has been adopted in 
QED on $S_1^n$ \cite{Adl,DS} and in 
the Yang-Mills theory on $S_1^n$ \cite{Sho} as a convenient gauge-fixing condition. 
(Adler proposed this condition in a study of  QED on $S_1^4$ \cite{Adl}.) 
The Adler condition can be generalized in such a manner that 
the generalized one contains a gauge parameter $\alpha$:  
\begin{align}
iL_{ab} \hat{A}_b +\alpha r_a \hat{B}=\hat{A}_a \,, 
\label{2.18}
\end{align}
where $\hat{B}$ is the Nakanishi-Lautrup field on $S_1^n$. 
This condition is expected to be useful for various studies of  
the Yang-Mills theory on $S_1^n$. 
However, the form of Eq. (\ref{2.18}) itself is not desirable  
for the ordinary first-order formalism of gauge fixing \cite{NL}, because Eq. (\ref{2.18})  
has an extra free index $a$ in comparison with the well-known (generalized) 
Lorentz condition $\partial_{\mu} A_{\mu}+\alpha B=0$. 
To avoid trouble with such an extra index, 
an alternative form of Eq. (\ref{2.18}),  
\begin{align}
ir_a L_{ab}\hat{A}_b +\alpha \hat{B}=0 \,, 
\label{2.19}
\end{align}
was considered in Ref. \cite{BD}.  
A simple way of observing the compatibility between 
Eqs. (\ref{2.18}) and (\ref{2.19}) is to contract Eq. (\ref{2.18}) by $r_{a}$. 
This immediately yields Eq. (\ref{2.19}) using Eqs. (\ref{2.0}) and (\ref{2.6}). 
(A complete proof of the equivalence between   
Eqs. (\ref{2.18}) and (\ref{2.19}) was given in Ref. \cite{BD}.)  
Equation (\ref{2.19}) is appropriate for  
the BRST gauge-fixing procedure \cite{KU,NO}.  
In fact, Eq. (\ref{2.19}) can be incorporated into the sum of gauge-fixing (GF) and  
Faddeev-Popov (FP) ghost terms 
(or simply the gauge-fixing term) 
written in the BRST-coboundary form
\begin{align}
S_{\rm GF}= \int d^{n} \varOmega \bigg\{ -i \boldsymbol\delta \mathrm{Tr}
\bigg[ \hat{\bar{C}} \bigg(ir_a L_{ab} \hat{A}_b +\frac{\alpha}{2} \hat{B} \bigg) 
\bigg] \bigg\} \,. 
\label{2.20}
\end{align}
The BRST transformation, denoted by $\boldsymbol\delta$, is defined by 
\begin{align}
&\boldsymbol\delta \hat{A}_a = ir_b \mathcal{L}_{ba} \hat{C} =P_{ab}\hat{D}_{b} \hat{C} \,,
\label{2.21}
\\
&\boldsymbol\delta \hat{C} =\frac{i}{2}\{ \hat{C}, \hat{C} \} \,,
\label{2.22}
\\
&\boldsymbol\delta \hat{\bar{C}} =i\hat{B}\,, 
\label{2.23}
\\
&\boldsymbol\delta \hat{B} =0 \,, 
\label{2.24}
\end{align}
where $\hat{C}$ and $\hat{\bar{C}}$ are the FP ghost and anti-ghost fields, 
respectively. 
The transformation rule (\ref{2.21}) is defined by replacing $\lambda$ in Eq. (\ref{2.7}) by 
$\hat{C}$. The nilpotency property $\boldsymbol{\delta}^{2}=0$ is valid for all the fields.  
In particular, $\boldsymbol{\delta}^{2}\hat{A}_{a}=0$ 
is verified by using the property $P_{ac}P_{cb}=P_{ab}$. The BRST invariance of $S_{\rm GF}$ 
is guaranteed by the nilpotency of $\boldsymbol{\delta}$, while that of $S_{\rm YM}$ is evident 
from its gauge invariance.

Carrying out the BRST transformation contained in the right-hand side of Eq.~(\ref{2.20}), 
we have
\begin{align}
S_{\rm GF}= \int d^{n} \varOmega\,  \mathrm{Tr}\bigg[ 
\hat{B} ir_a L_{ab} \hat{A}_b +\frac{\alpha}{2} \hat{B}^2 
-\frac{1}{2}i \hat{\bar{C}} L_{ab} \mathcal{L}_{ab} \hat{C}
\bigg] . 
\label{2.25}
\end{align}
Here the formula  
\begin{align}
r_a L_{ac}(r_b \mathcal{L}_{bc})=\frac{1}{2} L_{ab} \mathcal{L}_{ab} 
\label{2.26}
\end{align}
has been used. 
Integrating by parts over $(r_{\mu})$ and using Eqs. (\ref{2.0}), (\ref{2.4}) and 
(\ref{2.6}), we can rewrite Eq.~(\ref{2.25}) as 
\begin{align}
S_{\rm GF}& = \int d^{n} \varOmega\,  \mathrm{Tr}\bigg[ 
-(ir_b L_{ba} \hat{B}) \hat{A}_a +\frac{\alpha}{2} \hat{B}^2 
+ (ir_b L_{ba} \hat{\bar{C}}) r_{c}  \mathcal{L}_{ca} \hat{C}
\bigg] , 
\label{2.27}
\end{align}
where no existence of singularities of the fields has been assumed. 
Carrying out integration by parts again in Eq.~(\ref{2.27}) leads to   
\begin{align}
S_{\rm GF}= \int d^{n} \varOmega\, \mathrm{Tr}\bigg[ 
\hat{B} ir_a L_{ab} \hat{A}_b +\frac{\alpha}{2} \hat{B}^2 
-{1\over2}i (\mathcal{L}_{ab} L_{ab} \hat{\bar{C}}) \hat{C}
\bigg] 
\label{2.28}
\end{align}
by using the formula 
\begin{align}
r_a \mathcal{L}_{ac}(r_b L_{bc})=\frac{1}{2} \mathcal{L}_{ab} L_{ab} \,.
\label{2.29}
\end{align}

From the total action 
\begin{align}
S=S_{\rm YM}+S_{\rm GF} \,,
\label{2.30}
\end{align}
the Euler-Lagrange equations for $\hat{A}_a$, $\hat{B}$, $\hat{\bar{C}}$, and $\hat{C}$ are 
derived, respectively, as 
\begin{align}
&{i\over2} \mathcal{L}_{ab} \hat{F}_{abc}
=ir_b L_{bc} \hat{B}-\{ ir_b L_{bc} \hat{\bar{C}}, \hat{C} \} \,,
\label{2.31} 
\\ 
& ir_a L_{ab} \hat{A}_b +\alpha \hat{B}=0 \,, 
\label{2.32} 
\\
& L_{ab} \mathcal{L}_{ab} \hat{C}=0 \,, 
\label{2.33} 
\\
& \mathcal{L}_{ab} L_{ab} \hat{\bar{C}}=0 \,. 
\label{2.34}
\end{align}
Note here that the gauge-fixing condition (\ref{2.19}) has been obtained by varying 
$S_{\rm GF}$ with respect to $\hat{B}$. Applying $ir_{d} \mathcal{L}_{de}$ to Eq. (\ref{2.31})  
and contracting the indices $c$ and $e$ yield 
\begin{align}
\mathcal{L}_{ab} L_{ab} \hat{B}=\{ L_{ab} \hat{\bar{C}}, \mathcal{L}_{ab} \hat{C} \} 
\label{2.35}
\end{align}
after using Eqs. (\ref{2.0}), (\ref{2.29}) and (\ref{2.34}). 
The field equations (\ref{2.31})--(\ref{2.35}) 
have manifestly O($n+1$) covariant or invariant forms.
They are the spherical analogues of the field equations on the flat space presented in the literature 
\cite{KO,NO}.

Provided that $\alpha\neq 0$, we can eliminate the Nakanishi-Lautrup field $\hat{B}$ from 
Eq. (\ref{2.25}) using Eq. (\ref{2.32}) to obtain 
\begin{align}
S_{\rm GF}^{\prime}= \int d^{n} \varOmega\,  \mathrm{Tr}\bigg[ 
-\frac{1}{2\alpha} \big( ir_a L_{ab} \hat{A}_b \big)^2 
-\frac{1}{2}i \hat{\bar{C}} L_{ab} \mathcal{L}_{ab} \hat{C}
\bigg] . 
\label{2.36}
\end{align}
Carrying out integration by parts over $(r_{\mu})$ and using Eqs. (\ref{2.0}), (\ref{2.4}) and 
(\ref{2.6}), we can show that 
\begin{align}
&\int d^{n} \varOmega \big( ir_a L_{ab} \hat{A}_b \big)^2 
=-\int d^{n} \varOmega \hat{A}_{a} \partial_{a} \big( ir_b L_{bc} \hat{A}_c \big) 
\nonumber 
\\
&=-\int d^{n} \varOmega \hat{A}_{a} (L_{ac} L_{cb} +iL_{ab}) \hat{A}_{b} 
=-\int d^{n} \varOmega \hat{A}_{a} (L_{ac} L_{cb} +\delta_{ab}) \hat{A}_{b} \,.
\label{2.37}
\end{align}
Here, the identity 
\begin{align}
\hat{A}_{a} (iL_{ab} \hat{A}_{b} -\hat{A}_{a})=0 
\label{2.38}
\end{align}
has been used in the last equality. The identity (\ref{2.38}) is readily proven by using 
Eq. (\ref{2.6}) and its derivative with respect to $r_{\mu}$: 
$r_b \partial_a \hat{A}_{b}=-\hat{A}_{a} +(r_a/r_{n+1}) \hat{A}_{n+1}$ \cite{BD}.  
Substituting Eq. (\ref{2.37}) into Eq. (\ref{2.36}) leads to 
\begin{align}
S_{\rm GF}^{\prime}= \int d^{n} \varOmega\,  \mathrm{Tr}\bigg[ 
\frac{1}{2\alpha} \hat{A}_{a} (L_{ac} L_{cb} +\delta_{ab}) \hat{A}_{b} 
-\frac{1}{2}i \hat{\bar{C}} L_{ab} \mathcal{L}_{ab} \hat{C}
\bigg] . 
\label{2.39}
\end{align}
By using Eq. (\ref{2.38}), Eq. (\ref{2.39}) can be written 
\begin{align}
S_{\rm GF}^{\prime}
&= \int d^{n} \varOmega\,  \mathrm{Tr}\bigg[ 
\frac{1}{2\alpha} \hat{A}_{a} 
(L_{ac} +i\delta_{ac})\{L_{cb} +i(n-2)\delta_{cb}\} \hat{A}_{b} 
-\frac{1}{2}i \hat{\bar{C}} L_{ab} \mathcal{L}_{ab} \hat{C}
\bigg] 
\label{2.40}
\\
&= \int d^{n} \varOmega\,  \mathrm{Tr}\bigg[ 
\frac{1}{2\alpha} \hat{A}_{a} 
(L_{ac} +in\delta_{ac})(L_{cb} +i\delta_{cb}) \hat{A}_{b} 
-\frac{1}{2}i \hat{\bar{C}} L_{ab} \mathcal{L}_{ab} \hat{C}
\bigg] . 
\label{2.41}
\end{align}
Equations (\ref{2.40}) and (\ref{2.41}) 
are identical to the gauge-fixing terms adopted  
in earlier studies \cite{DS,Sho}. 


\section{\label{sec:level1}Supersphere}

In this section, we define a supersphere (or a supersymmetric sphere) 
and consider its associated symmetry group \cite{ND,susp}.

Let ${\bf R}^{n+1|2}$ be the $(n+1+2)$-dimensional Euclidian superspace with 
the Cartesian coordinate system $(\rho^A)=(\rho^a, \bar{\xi}, \xi)$ 
$(A=1,2,\ldots,n+1,-1,-2)$ that consists of commutative real numbers 
$(\rho^a)$ and anticommutative real numbers 
$(\rho^{-1}, \rho^{-2})\equiv(\bar{\xi}, \xi)$.  
(The complex conjugate of $\bar{\xi} \xi$ is defined by 
$(\bar{\xi} \xi)^{\ast}=\xi^{\ast} \bar{\xi}^{\ast}$ \cite{DeW}, 
so that $\bar{\xi} \xi$ is purely imaginary.) 
An $(n+2)$-dimensional unit supersphere $S_{1}^{n|2}$ 
embedded in ${\bf R}^{n+1|2}$ is defined by the constraint \cite{ND} 
\begin{align}
\rho^A g_{AB} \rho^B \equiv \rho^\mu \rho^\mu+(\rho^{n+1}){}^2 -2i\bar{\xi}\xi=1 
\,. 
\label{3.1}
\end{align}
Here $g_{AB}$ is a metric tensor on ${\bf R}^{n+1|2}$ whose non-vanishing components 
are
\begin{align}
g_{11}=g_{22}=\cdots=g_{(n+1)(n+1)}=1\,, \quad 
g_{-1-2}=-g_{-2-1}=-i \,.
\label{3.2}
\end{align}
The set of the linear transformations that leave $\rho^A g_{AB} \rho^B$ invariant 
forms the orthosymplectic supergroup ${\rm OSp}(n+1|2)$ \cite{FK,DeW}. 
By imposing the constraint (\ref{3.1}) to the superspace coordinates $(\rho^A)$,  
the ${\rm OSp}(n+1|2)$ symmetry that is linearly realized in ${\bf R}^{n+1|2}$  
is broken into the linear symmetry characterized by the subgroup ${\rm OSp}(n|2)$.  
In this sense, the supersphere $S_{1}^{n|2}$ can be 
represented as the coset superspace ${\rm OSp}(n+1|2)/{\rm OSp}(n|2)$ \cite{ND}. 
This is precisely a supersymmetric generalization of the coset space  
${\rm O}(n+1)/{\rm O}(n)$, which can be identified with the sphere $S_{1}^{n}$.  
The ${\rm OSp}(n+1|2)$ symmetry is realized on $S_{1}^{n|2}$ in a nonlinear way, 
as will be mentioned under Eq. (\ref{3.12}). 
Having imposed the constraint (\ref{3.1}) on $(\rho^A)$, we can use 
$(\rho^M)=(\rho^{\mu}, \bar{\xi}, \xi)$  $(M=1,2,\ldots,n,-1,-2)$ 
as local coordinates on $S_{1}^{n|2}$,  
treating $\rho^{n+1}=\pm(1-\rho^{\mu}\rho^{\mu}+2i\bar{\xi} \xi)^{1/2}$ as 
a dependent variable \footnotemark[4].  
%
\footnotetext[4]{The indices $A, B, C,$ and $D$ run from $-2$ to $n+1$ 
except for $0$, while the indices $M$ and $N$ run from $-2$ to $n$ except for $0$.}

We now rewrite this expression as 
\begin{align}  
\rho^{n+1}=\pm (1-\rho^{\mu} \rho^{\mu})^{1/2} 
\pm(1-\rho^{\mu} \rho^{\mu})^{-1/2} i\bar{\xi} \xi \,. 
\label{3} 
\end{align}
Thereby 
it becomes clear that because $\rho^{n+1}$ is purely real, the $\rho^{\mu} \rho^{\mu}$ 
has to be in the interval  $0\leq \rho^\mu \rho^\mu \leq 1$.  By virtue of this, 
it is  possible to embed the sphere $S_{1}^{n}$ in the supersphere $S_{1}^{n|2}$ 
by identifying $(\rho^{\mu})$ with the coordinates $(r_{\mu})$ on $S_{1}^{n}$ 
by simply setting $\rho^{\mu}= r_{\mu}$. (Recall here that $0\leq r_{\mu} r_{\mu} \leq 1$.) 
As a result, $S_{1}^{n}$ is considered to be a commutative subspace of $S_{1}^{n|2}$.  
It is easy to see that the dependent variables $\rho^{n+1}$ and 
$r_{n+1}=\pm\sqrt{1-r_{\mu} r_{\mu}}\,$ are related by 
\begin{align}
\rho^{n+1}=r_{n+1}-\frac{i\xi\bar{\xi}}{r_{n+1}} \,, 
\quad 
r_{n+1}=\rho^{n+1}+\frac{i\xi\bar{\xi}}{\rho^{n+1}} \,.
\label{3.3}
\end{align}
The relation $\rho^{\mu}= r_{\mu}$ and Eq. (\ref{3.3}) are 
brought together in the expressions    
\begin{align}
&\rho^{a}=r_{a}-\delta_{a(n+1)}\frac{i\xi\bar{\xi}}{r_{n+1}} \,, 
\label{3.4}
\\
& r_{a}=\rho^{a}+\delta_{a(n+1)}\frac{i\xi\bar{\xi}}{\rho^{n+1}} \,.
\label{3.5}
\end{align}

In terms of the coordinates $(\rho^M)$, the generators of ${\rm OSp}(n+1|2)$, 
denoted by $J_{AB}$, are represented as 
\begin{align}
&J_{MN}=-i\big{(} \rho_M \partial_N 
-(-1)^{|M||N|}
\rho_N \partial_M\big{)} \,, 
\quad \rho_M \equiv g_{MN} \rho^{N}, \,\;  
\partial_M \equiv \frac{\partial}{\partial \rho^{M}} \,,
\label{3.6}
\\ 
&J_{M (n+1)} =-J_{(n+1) M}=i\rho_{n+1}\partial_{M} \,, \quad 
\rho_{n+1}\equiv \rho^{n+1} \,,
\label{3.7}
\end{align}
or more concisely 
\begin{align}
J_{AB}=-i\big{(} \rho_A \partial_B 
-(-1)^{|A||B|}  \rho_B \partial_A\big{)}
\,, \quad \partial_A\equiv \delta_{A}{}^{M}\partial_{M} \,. 
\label{3.8}
\end{align}
Here, $|A|$ is a function of $A$, called its Grassmann parity, defined as  
$|A|=0$ for $A=1,2,\ldots, n+1$, and $|A|=1$ for $A=-1,-2$. 
The derivatives $\partial_{-1}$ and $\partial_{-2}$ denote left derivatives. 
The generators $J_{AB}$ can be expressed more concretely as 
\begin{subequations}
\label{3.9}
\begin{align}
&J_{ab} =-i( \rho_a \partial_b -\rho_b \partial_a ) \,, 
\label{3.9a}
\\ 
&J_{a\,-1} =-J_{-1a}=-i\rho_a \partial_{\bar{\xi}}+\xi \partial_{a} \,,  
\label{3.9b}
\\
&J_{a\,-2} =-J_{-2a}=-i\rho_a \partial_{\xi}-\bar{\xi} \partial_{a} \,, 
\label{3.9c}
\\
&J_{-1-1} =-2\xi\partial_{\bar{\xi}} \,, 
\quad
J_{-2-2} =2\bar{\xi}\partial_{\xi} \,, 
\label{3.9d}
\\
&J_{-1-2} =J_{-2-1} 
=-\xi \partial_{\xi}+ \bar{\xi} \partial_{\bar{\xi}} \,, 
\label{3.9e}
\end{align}
\end{subequations}
where 
$\partial_{a}\equiv \delta_a{}^{\mu} \partial/\partial \rho^{\mu}$,  
$\partial_{\bar{\xi}}\equiv \partial/\partial\bar{\xi}\,(=\partial_{-1}),$ and  
$\partial_{\xi}\equiv \partial/\partial\xi\,(=\partial_{-2})$.   
Using Eq. (\ref{3.4}), $J_{ab}$ can be written 
\begin{align}
J_{ab}=L_{ab}-\frac{\xi\bar{\xi}}{r_{n+1}}
(\delta_{a(n+1)}\partial_{b}-\delta_{b(n+1)}\partial_{a}) \,,
\label{3.10}
\end{align}
with $L_{ab}$ in Eq. (\ref{2.3}). 
Note here that $J_{\mu\nu}$ is equal to $L_{\mu\nu}$, 
whereas $J_{\mu(n+1)}$ is not equal to $L_{\mu(n+1)}$. Using 
\begin{align}
\frac{\partial \rho_{n+1}}{\partial \rho^{M}}=-\frac{\rho_{M}}{\rho_{n+1}} \,, 
\label{3.11}
\end{align}
we can show that the generators in Eqs. (\ref{3.6}) and (\ref{3.7}) satisfy the 
supercommutation relations of the OSp($n+1/2$) super Lie algebra,  
\begin{align}
&[J_{AB}, J_{CD} \} \equiv 
J_{AB} J_{CD} 
-(-1)^{(|A|+|B|)(|C|+|D|)} J_{CD} J_{AB}  
\nonumber \\ 
&= i \big( (-1)^{(|B|+1)|C|} g_{AC} J_{BD} 
-(-1)^{|A||B|+(|A|+1)|C|} g_{BC} J_{AD} 
\nonumber \\ 
&\qquad -(-1)^{(|B|+|C|+1) |D|} g_{AD} J_{BC}
+(-1)^{|A||B|+(|A|+|C|+1)|D|} g_{BD} J_{AC} \big) \,.
\label{3.12}
\end{align}
Thus, the OSp($n+1/2$) symmetry is realized on $S_1^{n|2}$.  
Because $J_{M (n+1)}$ is nonlinear with respect to $\rho^{N}$,  
one says that this symmetry is nonlinearly realized.  
It is now clear that the generators $\{J_{ab}\}$ generate 
the Lie group O($n+1$), while the generators 
$J_{-1-1}$, $J_{-1-2}$, and $J_{-2-2}$ generate the Lie group Sp($2$).  
The remainders $J_{a\,-1}$ and $J_{a\,-2}$ 
are generators of the rotational supersymmetry.


\section{\label{sec:level1}Yang-Mills superfield on supersphere}

This section treats a Yang-Mills field on the supersphere $S^{n|2}_1$ 
and its component fields. Because the Yang-Mills field on $S^{n|2}_1$  
is a superfield, we shall refer to it as the {\em Yang-Mills superfield}. 
The tensor components of the rank-3 field strength tensor of the Yang-Mills superfield 
will be written in terms of the component fields.

Let $\hat{\mathcal{A}}_{A}$ be a (Hermitian) Yang-Mills superfield on $S^{n|2}_{1}$ 
that takes values in the Lie algebra ${\frak g}$.  
Then, similarly to the Yang-Mills field $\hat{A}_{a}$ on $S^{n}_{1}$, 
the superfield $\hat{\mathcal{A}}_{A}$ can be expanded as 
$\hat{\mathcal{A}}_{A}=\hat{\mathcal{A}}_{A}^{i}T^{i}$. 
Because $\hat{\mathcal{A}}_{a}$ is associated with $\partial_a$, it is a {\em commutative}  
superfield, while because $\hat{\mathcal{A}}_{-1}$ and $\hat{\mathcal{A}}_{-2}$ 
are associated with $\partial_{-1}$ and $\partial_{-2}$, respectively, they are 
{\em anticommutative} superfields. 
The Yang-Mills superfield $\hat{\mathcal{A}}_{A}=\hat{\mathcal{A}}_{A}(\rho^M)$ 
can be expanded about $\bar{\xi}=\xi=0$ in the sense of the Taylor expansion: 
\begin{align}
\hat{\mathcal{A}}_{A}(\rho^\mu, \bar{\xi}, \xi)
=\hat{\mathcal{A}}_{A}(\rho^\mu, 0,0)
+\bar{\xi}(\partial_{\bar{\xi}} \hat{\mathcal{A}}_{A})_0 
+\xi(\partial_{\xi} \hat{\mathcal{A}}_{A})_0 
+\xi\bar{\xi} (\partial^2 \hat{\mathcal{A}}_{A})_0 \,,
\label{4.1}
\end{align}
where
\begin{subequations}
\label{4.2}
\begin{align}
(\partial_{\bar{\xi}} \hat{\mathcal{A}}_{A})_0 
& \equiv \frac{\partial\hat{\mathcal{A}}_{A}(\rho^M)}{\partial\bar{\xi}}
\bigg\vert_{\bar{\xi}=\xi=0} \,, 
\label{4.2a}  
\\
(\partial_{\xi} \hat{\mathcal{A}}_{A})_0 
& \equiv \frac{\partial\hat{\mathcal{A}}_{A}(\rho^M)}{\partial\xi}
\bigg\vert_{\bar{\xi}=\xi=0} \,, 
\label{4.2b}
\\ 
(\partial^2 \hat{\mathcal{A}}_{A})_0 
& \equiv \frac{\partial^2 \hat{\mathcal{A}}_{A}(\rho^M)}
{\partial\bar{\xi}\partial\xi}
\bigg\vert_{\bar{\xi}=\xi=0} \,. 
\label{4.2c}
\end{align}
\end{subequations}
All the expansion coefficients in Eq. (\ref{4.1}) are functions of $(\rho^\mu)$ 
and may be refered to as the {\em component fields} of 
$\hat{\mathcal{A}}_{A}(\rho^M)$. 
Because $\rho^{\mu}=r_{\mu}$, these fields are functions of $(r_{\mu})$, and hence 
they are regarded as fields on $S_1^n$.   
As will be confirmed later,  
the vector field $\hat{\mathcal{A}}_{a}(\rho^\mu, 0,0)$ is identified with 
the Yang-Mills field $\hat{A}_{a}$, while $\hat{\mathcal{A}}_{-1}(\rho^\mu, 0,0)$ 
and $\hat{\mathcal{A}}_{-2}(\rho^\mu, 0,0)$ are identified with 
the FP ghost field $\hat{C}$ and 
the FP anti-ghost field $\hat{\bar{C}}$, respectively, 
up to the imaginary unit $i$: 
\begin{align}
\hat{A}_{a}(r_\mu) \equiv\hat{\mathcal{A}}_{a}(\rho^\mu, 0,0) \,, 
\quad 
\hat{C}(r_\mu) \equiv i\hat{\mathcal{A}}_{-1}(\rho^\mu, 0,0) \,, 
\quad 
\hat{\bar{C}}(r_\mu) \equiv i\hat{\mathcal{A}}_{-2}(\rho^\mu, 0,0) \,.
\label{4.3}
\end{align}
The imaginary unit in Eq. (\ref{4.3}) is necessary so that  
$\hat{C}$ and $\hat{\bar{C}}$ can be purely real. 
With this identification, 
it is desirable for later discussions to express Eq. (\ref{4.1}) as  
\begin{subequations}
\label{4.4}
\begin{align}
&\hat{\mathcal{A}}_{a}= \hat{A}_{a} 
+\bar{\xi}(\partial_{\bar{\xi}} \hat{\mathcal{A}}_{a})_0 
+\xi(\partial_{\xi} \hat{\mathcal{A}}_{a})_0 
+\xi\bar{\xi} (\partial^2 \hat{\mathcal{A}}_{a})_0 \,, 
\label{4.4a}
\\ 
&\hat{\mathcal{A}}_{-1}= -i\hat{C} 
+\bar{\xi}(\partial_{\bar{\xi}} \hat{\mathcal{A}}_{-1})_0 
+\xi(\partial_{\xi} \hat{\mathcal{A}}_{-1})_0 
+\xi\bar{\xi} (\partial^2 \hat{\mathcal{A}}_{-1})_0 \,, 
\label{4.4b}
\\ 
&\hat{\mathcal{A}}_{-2}= -i\hat{\bar{C}} 
+\bar{\xi}(\partial_{\bar{\xi}} \hat{\mathcal{A}}_{-2})_0 
+\xi(\partial_{\xi} \hat{\mathcal{A}}_{-2})_0 
+\xi\bar{\xi} (\partial^2 \hat{\mathcal{A}}_{-2})_0 \,.
\label{4.4c}
\end{align}
\end{subequations}
Note here that $\hat{A}_{a}$, $(\partial^2 \hat{\mathcal{A}}_{a})_0$, 
$(\partial_{\bar{\xi}} \hat{\mathcal{A}}_{-1})_0$, 
$(\partial_{\xi} \hat{\mathcal{A}}_{-1})_0$, 
$(\partial_{\bar{\xi}} \hat{\mathcal{A}}_{-2})_0$, and 
$(\partial_{\xi} \hat{\mathcal{A}}_{-2})_0$ are 
{\em commutative} component fields, 
while $(\partial_{\bar{\xi}} \hat{\mathcal{A}}_{a})_0$, 
$(\partial_{\xi} \hat{\mathcal{A}}_{a})_0$, $\hat{C}$, 
$(\partial^2 \hat{\mathcal{A}}_{-1})_0$, $\hat{\bar{C}}$, and 
$(\partial^2 \hat{\mathcal{A}}_{-2})_0$ are 
{\em anticommutative} component fields.

The Yang-Mills superfield $\hat{\mathcal{A}}_{A}$ is assumed to live on 
the tangent superspace, $\mathfrak{T}_P S_1^{n|2}$, at a point 
$P(\rho^{M})$ on $S_1^{n|2}$ by imposing the transversality 
condition \footnotemark[5] 
%
\footnotetext[5]{As usual, we assign the ghost numbers $0$, $1$ and $-1$ to 
$\hat{A}_{a}$, $\hat{C}$, and $\hat{\bar{C}}$, respectively \cite{KO,NO}.   
Then, it is natural to assign the ghost numbers $0$, $1$, and $-1$ also to 
$\rho^{a}$, $\xi$, and $\bar{\xi}$, respectively, in such a way that 
$\rho^{A} \hat{\mathcal{A}}_{A}$ has the definite ghost number 0.}
%
%
\begin{align}
\rho^{A} \hat{\mathcal{A}}_{A}
=\rho^a \hat{\mathcal{A}}_{a}+\bar{\xi} \hat{\mathcal{A}}_{-1}
+\xi\hat{\mathcal{A}}_{-2} =0 \,.
\label{4.5}
\end{align}
This is precisely a supersymmetric analogue of Eq. (\ref{2.6}). 
Substituting Eqs. (4.4) into Eq. (\ref{4.5}) and 
using Eq. (\ref{3.4}), we have from each power in $(\bar{\xi}, \xi)$,  
\begin{align}
& r_a \hat{A}_a =0 \,, 
\label{4.6} 
\\ 
& r_a (\partial_{\bar{\xi}} \hat{\mathcal{A}}_{a})_{0}
-i\hat{C} =0 \,, 
\label{4.7} 
\\
& r_a (\partial_{\xi} \hat{\mathcal{A}}_{a})_{0}
-i\hat{\bar{C}} =0 \,, 
\label{4.8}
\\ 
& r_a (\partial^2 \hat{\mathcal{A}}_{a})_{0} 
-\frac{i}{r_{n+1}} \hat{A}_{n+1} 
-(\partial_{\xi} \hat{\mathcal{A}}_{-1})_{0}
+(\partial_{\bar{\xi}} \hat{\mathcal{A}}_{-2})_{0}=0 \,. 
\label{4.9}
\end{align}
Equation (\ref{4.6}) is nothing but the transversality condition 
(\ref{2.6}), while Eqs. (\ref{4.7})--(\ref{4.9}) are recognized  
to be new conditions associated with Eq. (\ref{4.6}).  
Equations (\ref{4.7})--(\ref{4.9}) imply that the fields 
$(\partial_{\bar{\xi}} \hat{\mathcal{A}}_{a})_{0}$, 
$(\partial_{\xi} \hat{\mathcal{A}}_{a})_{0}$, and 
$(\partial^2 \hat{\mathcal{A}}_{a})_{0}$ live outside the tangent space 
$\mathfrak{T}_P S_1^{n}$. 
Their normal (or radial) components are completely determined by 
Eqs. (\ref{4.7})--(\ref{4.9}).

As a generalization of Eq. (\ref{2.11a}), 
the field strength of $\hat{\mathcal{A}}_{A}$ is given in     
a manifestly OSp($n+1/2$)-covariant form, 
\begin{align}
\hat{\mathcal{F}}_{ABC}=\;& i\big( J_{AB}\hat{\mathcal{A}}_{C}
-\rho_A [\hat{\mathcal{A}}_{B}, \hat{\mathcal{A}}_{C} \} \big)
\nonumber \\ 
& +(-1)^{|A|(|B|+|C|)} i\big( J_{BC}\hat{\mathcal{A}}_{A}
-\rho_B [\hat{\mathcal{A}}_{C}, \hat{\mathcal{A}}_{A} \} \big) 
\nonumber \\
& +(-1)^{|C|(|A|+|B|)} i\big( J_{CA}\hat{\mathcal{A}}_{B}
-\rho_C [\hat{\mathcal{A}}_{A}, \hat{\mathcal{A}}_{B} \} \big) \,,
\label{4.10}
\end{align}
where the supercommutation relation 
$[\hat{\mathcal{A}}_{A}, \hat{\mathcal{A}}_{B} \}$ is understood as \cite{DeW} 
\begin{align}
[\hat{\mathcal{A}}_{A}, \hat{\mathcal{A}}_{B} \} 
\equiv 
\hat{\mathcal{A}}_{A}\hat{\mathcal{A}}_{B}
-(-1)^{|A||B|} \hat{\mathcal{A}}_{B}\hat{\mathcal{A}}_{A} \,. 
\label{4.11}
\end{align}
This obeys the generalized antisymmetric rule 
$[\hat{\mathcal{A}}_{A}, \hat{\mathcal{A}}_{B} \} 
=-(-1)^{|A||B|} [\hat{\mathcal{A}}_{B}, \hat{\mathcal{A}}_{A} \}$. 
Because $J_{AB}$ obeys the same rule, $J_{AB}=-(-1)^{|A||B|}J_{BA}$, 
it is easy to see that the field strength $\hat{\mathcal{F}}_{ABC}$ has   
the generalized totally-antisymmetric property    
\begin{align}
\hat{\mathcal{F}}_{ABC} =-(-1)^{|A||B|}\hat{\mathcal{F}}_{BAC} 
=(-1)^{|A|(|B|+|C|)} \hat{\mathcal{F}}_{BCA} \,. 
\label{4.12}
\end{align}

Substituting Eqs. (\ref{3.9a}) and (\ref{4.4a}) into Eq. (\ref{4.10}) with    
$A=a$, $B=b$, $C=c$, and using Eqs. (\ref{3.4}) and (\ref{3.10}) to express 
$\hat{\mathcal{F}}_{abc}$ as a power series in $(\bar{\xi}, \xi)$, we obtain 
\begin{align}
\hat{\mathcal{F}}_{abc}&= i( J_{ab}\hat{\mathcal{A}}_{c}
-\rho_a [\hat{\mathcal{A}}_{b}, \hat{\mathcal{A}}_{c} ] \big) 
+\mbox{cyclic permutations in $(a, b, c)$}
\nonumber \\ 
&= \hat{F}_{abc} 
+i\bar{\xi} \big( \mathcal{L}_{ab}(\partial_{\bar{\xi}} \hat{\mathcal{A}}_{c})_0 
+\mbox{c.p.} \big) 
+i\xi \big( \mathcal{L}_{ab}(\partial_{\xi} \hat{\mathcal{A}}_{c})_0 
+\mbox{c.p.} \big) 
\nonumber \\ 
&\;\quad +i\xi\bar{\xi} \bigg( -\frac{1}{r_{n+1}} \hat{F}_{ab} \delta_{c(n+1)} 
+\mathcal{L}_{ab}(\partial^{2} \hat{\mathcal{A}}_{c})_0  
\nonumber \\ 
&\;\quad -r_{a} \{ (\partial_{\bar{\xi}} \hat{\mathcal{A}}_{b})_0, 
(\partial_\xi \hat{\mathcal{A}}_{c})_0 \} 
+r_{a} \{ (\partial_{\bar{\xi}} \hat{\mathcal{A}}_{c})_0, 
(\partial_{\xi} \hat{\mathcal{A}}_{b})_0 \} 
+\mbox{c.p.} \bigg) \,, 
\label{4.13}
\end{align}
where $\mathcal{L}_{ab}$, $\hat{F}_{abc}$, and $\hat{F}_{ab}$ 
are given in Eqs. (\ref{2.8}),  (\ref{2.11}), and (\ref{2.12}), respectively. 
Substituting some of  Eqs. (\ref{3.9a})--(\ref{3.9c}), and (4.4) into 
Eq. (\ref{4.10}) with $A=a$, $B=b$, $C=-1, -2$, 
and using Eqs. (\ref{3.4}) and (\ref{3.10}), we obtain
\begin{align}
\hat{\mathcal{F}}_{ab\,-1}&
=i(J_{ab}\hat{\mathcal{A}}_{-1}-[\rho_{a} \hat{\mathcal{A}}_{b}
-\rho_{b} \hat{\mathcal{A}}_{a}, \hat{\mathcal{A}}_{-1}] 
-J_{a\,-1}\hat{\mathcal{A}}_{b} +J_{b\,-1}\hat{\mathcal{A}}_{a} 
-\rho_{-1} [ \hat{\mathcal{A}}_{a}, \hat{\mathcal{A}}_{b} ] )
\nonumber \\ 
&=\mathcal{L}_{ab}\hat{C}
-r_{a}(\partial_{\bar{\xi}} \hat{\mathcal{A}}_{b})_0 
+r_{b}(\partial_{\bar{\xi}} \hat{\mathcal{A}}_{a})_0 
\nonumber \\ 
&\;\quad 
+i\bar{\xi} \big( \mathcal{L}_{ab}(\partial_{\bar{\xi}} \hat{\mathcal{A}}_{-1})_0 
+i \{ \hat{C},\, r_{a}(\partial_{\bar{\xi}} \hat{\mathcal{A}}_{b})_0 
-r_{b}(\partial_{\bar{\xi}} \hat{\mathcal{A}}_{a})_0 \} \big)
\nonumber \\ 
&\;\quad 
+i\xi \big( \mathcal{L}_{ab}(\partial_{\xi} \hat{\mathcal{A}}_{-1})_0 
+i \{ \hat{C},\, r_{a}(\partial_{\xi} \hat{\mathcal{A}}_{b})_0 
-r_{b}(\partial_{\xi} \hat{\mathcal{A}}_{a})_0 \} 
\nonumber \\ 
&\;\quad
-\hat{F}_{ab} -ir_{a}(\partial^{2} \hat{\mathcal{A}}_{b})_0
+ir_{b}(\partial^{2} \hat{\mathcal{A}}_{a})_0 \big)
\nonumber \\ 
&\;\quad +i\xi\bar{\xi} \bigg( \frac{1}{r_{n+1}} \big\{ \delta_{a(n+1)} 
\big( i\hat{D}_b \hat{C} +(\partial_{\bar{\xi}} \hat{\mathcal{A}}_{b})_0 \big) 
-\delta_{b(n+1)} 
\big( i\hat{D}_a \hat{C} +(\partial_{\bar{\xi}} \hat{\mathcal{A}}_{a})_0 \big) 
\big\}
\nonumber \\ 
&\;\quad
+\mathcal{L}_{ab} (\partial^{2} \hat{\mathcal{A}}_{-1})_0
-\hat{D}_a (\partial_{\bar{\xi}} \hat{\mathcal{A}}_{b})_0 
+\hat{D}_b (\partial_{\bar{\xi}} \hat{\mathcal{A}}_{a})_0 
\nonumber \\ 
&\;\quad
-i[ \hat{C},\, r_{a}(\partial^2 \hat{\mathcal{A}}_{b})_0 
-r_{b}(\partial^2 \hat{\mathcal{A}}_{a})_0 ] 
\nonumber \\ 
&\;\quad
-[ (\partial_{\bar{\xi}} \hat{\mathcal{A}}_{-1})_0, \,
r_{a}(\partial_{\xi} \hat{\mathcal{A}}_{b})_0 
-r_{b}(\partial_{\xi} \hat{\mathcal{A}}_{a})_0 ] 
\nonumber \\ 
&\;\quad
+[ (\partial_{\xi} \hat{\mathcal{A}}_{-1})_0, \,
r_{a}(\partial_{\bar{\xi}} \hat{\mathcal{A}}_{b})_0 
-r_{b}(\partial_{\bar{\xi}} \hat{\mathcal{A}}_{a})_0 ]
\bigg) \,, 
\label{4.14}
\\
\hat{\mathcal{F}}_{ab\,-2}&
=i(J_{ab}\hat{\mathcal{A}}_{-2}-[\rho_{a} \hat{\mathcal{A}}_{b}
-\rho_{b} \hat{\mathcal{A}}_{a}, \hat{\mathcal{A}}_{-2}] 
-J_{a\,-2}\hat{\mathcal{A}}_{b} +J_{b\,-2}\hat{\mathcal{A}}_{a} 
-\rho_{-2} [ \hat{\mathcal{A}}_{a}, \hat{\mathcal{A}}_{b} ] )
\nonumber \\ 
&=\mathcal{L}_{ab}\hat{\bar{C}}
-r_{a}(\partial_{\xi} \hat{\mathcal{A}}_{b})_0 
+r_{b}(\partial_{\xi} \hat{\mathcal{A}}_{a})_0 
\nonumber \\ 
&\;\quad 
+i\bar{\xi} \big( \mathcal{L}_{ab}(\partial_{\bar{\xi}} \hat{\mathcal{A}}_{-2})_0 
+i \{ \hat{\bar{C}},\, r_{a}(\partial_{\bar{\xi}} \hat{\mathcal{A}}_{b})_0 
-r_{b}(\partial_{\bar{\xi}} \hat{\mathcal{A}}_{a})_0 \} 
\nonumber \\ 
&\;\quad
+\hat{F}_{ab} +ir_{a}(\partial^{2} \hat{\mathcal{A}}_{b})_0
-ir_{b}(\partial^{2} \hat{\mathcal{A}}_{a})_0 \big)
\nonumber \\ 
&\;\quad 
+i\xi \big( \mathcal{L}_{ab}(\partial_{\xi} \hat{\mathcal{A}}_{-2})_0 
+i \{ \hat{\bar{C}},\, r_{a}(\partial_{\xi} \hat{\mathcal{A}}_{b})_0 
-r_{b}(\partial_{\xi} \hat{\mathcal{A}}_{a})_0 \} \big) 
\nonumber \\ 
&\;\quad
+i\xi\bar{\xi} \bigg( \frac{1}{r_{n+1}} \big\{ \delta_{a(n+1)} 
\big( i\hat{D}_b \hat{\bar{C}} +(\partial_{\xi} \hat{\mathcal{A}}_{b})_0 \big) 
-\delta_{b(n+1)} 
\big( i\hat{D}_a \hat{\bar{C}} +(\partial_{\xi} \hat{\mathcal{A}}_{a})_0 \big) 
\big\}
\nonumber \\ 
&\;\quad
+\mathcal{L}_{ab} (\partial^{2} \hat{\mathcal{A}}_{-2})_0
-\hat{D}_a (\partial_{\xi} \hat{\mathcal{A}}_{b})_0 
+\hat{D}_b (\partial_{\xi} \hat{\mathcal{A}}_{a})_0 
\nonumber \\ 
&\;\quad
-i[ \hat{\bar{C}},\, r_{a}(\partial^2 \hat{\mathcal{A}}_{b})_0 
-r_{b}(\partial^2 \hat{\mathcal{A}}_{a})_0 ] 
\nonumber \\ 
&\;\quad
-[ (\partial_{\bar{\xi}} \hat{\mathcal{A}}_{-2})_0, \,
r_{a}(\partial_{\xi} \hat{\mathcal{A}}_{b})_0 
-r_{b}(\partial_{\xi} \hat{\mathcal{A}}_{a})_0 ] 
\nonumber \\ 
&\;\quad
+[ (\partial_{\xi} \hat{\mathcal{A}}_{-2})_0, \,
r_{a}(\partial_{\bar{\xi}} \hat{\mathcal{A}}_{b})_0 
-r_{b}(\partial_{\bar{\xi}} \hat{\mathcal{A}}_{a})_0 ]
\bigg) \,, 
\label{4.15}
\end{align}
where $\hat{D}_a$ is the covariant derivative given in Eqs. (\ref{2.10}).  
In the same manner, we also obtain from  
Eqs. (\ref{3.9b}), (\ref{3.9c}), (\ref{3.9d}), (\ref{3.9e}), 
(4.4), and (\ref{4.10}), 
\begin{align}
\hat{\mathcal{F}}_{a\,-1-1}&= i( 2J_{a\,-1}\hat{\mathcal{A}}_{-1}
+2r_{-1}[\hat{\mathcal{A}}_{a}, \hat{\mathcal{A}}_{-1}]
+J_{-1-1}\hat{\mathcal{A}}_{a}
-r_{a}\{\hat{\mathcal{A}}_{-1}, \hat{\mathcal{A}}_{-1}\} )
\nonumber \\ 
&= r_a \big( 2(\partial_{\bar{\xi}} \hat{\mathcal{A}}_{-1})_0 
+i\{ \hat{C}, \hat{C} \} \big)
\nonumber \\ 
&\;\quad
+2\bar{\xi} r_a [\hat{C}, (\partial_{\bar{\xi}} \hat{\mathcal{A}}_{-1})_0 ]
\nonumber \\ 
&\;\quad
+2\xi \big(\hat{D}_a \hat{C} -i(\partial_{\bar{\xi}} \hat{\mathcal{A}}_{a})_0
-r_a (\partial^2 \hat{\mathcal{A}}_{-1})_0
+r_a [\hat{C}, (\partial_{\xi} \hat{\mathcal{A}}_{-1})_0 ]\big) 
\nonumber \\ 
&\;\quad
+2i\xi\bar{\xi} \big(\hat{D}_a (\partial_{\bar{\xi}} \hat{\mathcal{A}}_{-1})_0 
-\{ \hat{C}, (\partial_{\bar{\xi}} \hat{\mathcal{A}}_{a})_0 \}
+ir_a \{ \hat{C}, (\partial^2 \hat{\mathcal{A}}_{-1})_0 \} 
\nonumber \\ 
&\;\quad
+r_a [ (\partial_{\bar{\xi}} \hat{\mathcal{A}}_{-1})_0, 
(\partial_{\xi} \hat{\mathcal{A}}_{-1})_0 ] \big)
\,, 
\label{4.16}
\\
\hat{\mathcal{F}}_{a\,-1-2}&
= i( J_{a\,-1}\hat{\mathcal{A}}_{-2} 
+r_{-1}[\hat{\mathcal{A}}_{a}, \hat{\mathcal{A}}_{-2}] 
+J_{a\,-2}\hat{\mathcal{A}}_{-1}
+r_{-2}[\hat{\mathcal{A}}_{a}, \hat{\mathcal{A}}_{-1}]
\nonumber \\ 
&\;\quad
+J_{-1-2}\hat{\mathcal{A}}_{a}
-r_{a}\{\hat{\mathcal{A}}_{-1}, \hat{\mathcal{A}}_{-2}\} )
\nonumber \\ 
&= r_a \big( (\partial_{\bar{\xi}} \hat{\mathcal{A}}_{-2})_0 
+(\partial_{\xi} \hat{\mathcal{A}}_{-1})_0
+i\{ \hat{C}, \hat{\bar{C}} \} \big)
\nonumber \\
&\;\quad
-\bar{\xi} \big(\hat{D}_a \hat{C} -i(\partial_{\bar{\xi}} \hat{\mathcal{A}}_{a})_0 
-r_a (\partial^2 \hat{\mathcal{A}}_{-1})_0
-r_a [\hat{C}, (\partial_{\bar{\xi}} \hat{\mathcal{A}}_{-2})_0 ]
\nonumber \\ 
&\;\quad
-r_a [\hat{\bar{C}}, (\partial_{\bar{\xi}} \hat{\mathcal{A}}_{-1})_0 ]
\big) 
\nonumber \\ 
&\;\quad
+\xi \big(\hat{D}_a \hat{\bar{C}} -i(\partial_{\xi} \hat{\mathcal{A}}_{a})_0 
-r_a (\partial^2 \hat{\mathcal{A}}_{-2})_0
+r_a [\hat{C}, (\partial_{\xi} \hat{\mathcal{A}}_{-2})_0 ]
\nonumber \\ 
&\;\quad
+r_a [\hat{\bar{C}}, (\partial_{\xi} \hat{\mathcal{A}}_{-1})_0 ]
\big) 
\nonumber \\ 
&\;\quad
+i\xi\bar{\xi} \big(\hat{D}_a (\partial_{\bar{\xi}} \hat{\mathcal{A}}_{-2})_0 
+\hat{D}_a (\partial_{\xi} \hat{\mathcal{A}}_{-1})_0
-\{ \hat{C}, (\partial_{\xi} \hat{\mathcal{A}}_{a})_0 \}
-\{ \hat{\bar{C}}, (\partial_{\bar{\xi}} \hat{\mathcal{A}}_{a})_0 \}
\nonumber \\ 
&\;\quad
+ir_a \{ \hat{C}, (\partial^2 \hat{\mathcal{A}}_{-2})_0 \} 
+ir_a \{ \hat{\bar{C}}, (\partial^2 \hat{\mathcal{A}}_{-1})_0 \} 
+r_a [ (\partial_{\bar{\xi}} \hat{\mathcal{A}}_{-1})_0, 
(\partial_{\xi} \hat{\mathcal{A}}_{-2})_0 ] 
\nonumber \\ 
&\;\quad
-r_a [ (\partial_{\xi} \hat{\mathcal{A}}_{-1})_0, 
(\partial_{\bar{\xi}} \hat{\mathcal{A}}_{-2})_0 ] \big) \,, 
\label{4.17} 
\\
\hat{\mathcal{F}}_{a\,-2-2}&= 
i(2J_{a\,-2}\hat{\mathcal{A}}_{-2}
+2r_{-2}[\hat{\mathcal{A}}_{a}, \hat{\mathcal{A}}_{-2}]
+J_{-2-2}\hat{\mathcal{A}}_{a}
-r_{a}\{\hat{\mathcal{A}}_{-2}, \hat{\mathcal{A}}_{-2}\} )
\nonumber \\ 
&= r_a \big( 2(\partial_{\xi} \hat{\mathcal{A}}_{-2})_0 
+i\{ \hat{\bar{C}}, \hat{\bar{C}} \} \big)
\nonumber \\ 
&\;\quad
-2\bar{\xi} \big(\hat{D}_a \hat{\bar{C}} 
-i(\partial_{\xi} \hat{\mathcal{A}}_{a})_0
-r_a (\partial^2 \hat{\mathcal{A}}_{-2})_0
-r_a [\hat{\bar{C}}, (\partial_{\bar{\xi}} \hat{\mathcal{A}}_{-2})_0 ]\big)
\nonumber \\ 
&\;\quad
+2\xi r_a [\hat{\bar{C}}, (\partial_{\xi} \hat{\mathcal{A}}_{-2})_0 ]
\nonumber \\ 
&\;\quad
+2i\xi\bar{\xi} \big(\hat{D}_a (\partial_{\xi} \hat{\mathcal{A}}_{-2})_0 
-\{ \hat{\bar{C}}, (\partial_{\xi} \hat{\mathcal{A}}_{a})_0 \}
+ir_a \{ \hat{\bar{C}}, (\partial^2 \hat{\mathcal{A}}_{-2})_0 \} 
\nonumber \\ 
&\;\quad
+r_a [ (\partial_{\bar{\xi}} \hat{\mathcal{A}}_{-2})_0, 
(\partial_{\xi} \hat{\mathcal{A}}_{-2})_0 ] \big)
\,, 
\label{4.18}
\end{align}
\begin{align}
\hat{\mathcal{F}}_{-1-1-1}&=
3i( J_{-1-1}\hat{\mathcal{A}}_{-1}
-r_{-1}\{\hat{\mathcal{A}}_{-1}, \hat{\mathcal{A}}_{-1}\} ) 
\nonumber \\ 
& =- 3i \xi \big( 2(\partial_{\bar{\xi}} \hat{\mathcal{A}}_{-1})_0 
+i\{ \hat{C}, \hat{C} \} \big) 
-6i\xi\bar{\xi}\, [\hat{C}, (\partial_{\bar{\xi}} \hat{\mathcal{A}}_{-1})_0 ]
\,, 
\label{4.19}
\\ 
\hat{\mathcal{F}}_{-1-1-2}&=
i( J_{-1-1}\hat{\mathcal{A}}_{-2}
-r_{-2}\{\hat{\mathcal{A}}_{-1}, \hat{\mathcal{A}}_{-1}\} 
+2J_{-1-2}\hat{\mathcal{A}}_{-1}
-2r_{-1}\{\hat{\mathcal{A}}_{-1}, \hat{\mathcal{A}}_{-2}\} ) 
\nonumber \\ 
& 
=- 2i \xi \big( (\partial_{\bar{\xi}} \hat{\mathcal{A}}_{-2})_0 
+(\partial_{\xi} \hat{\mathcal{A}}_{-1})_0
+i\{ \hat{C}, \hat{\bar{C}} \} \big) 
\nonumber \\ 
&\;\quad
+i\bar{\xi} \big( 2(\partial_{\bar{\xi}} \hat{\mathcal{A}}_{-1})_0 
+i\{ \hat{C}, \hat{C} \} \big) 
\nonumber \\ 
&\;\quad
-2i\xi\bar{\xi} \big( 
[ \hat{C},\, (\partial_{\bar{\xi}} \hat{\mathcal{A}}_{-2})_0 
+(\partial_{\xi} \hat{\mathcal{A}}_{-1})_0 ] 
+[ \hat{\bar{C}}, (\partial_{\bar{\xi}} \hat{\mathcal{A}}_{-1})_0 ] \big) 
\,, 
\label{4.20}
\\ 
\hat{\mathcal{F}}_{-1-2-2}&=
i( 2J_{-1-2}\hat{\mathcal{A}}_{-2}
-2r_{-2}\{\hat{\mathcal{A}}_{-1}, \hat{\mathcal{A}}_{-2}\}
+J_{-2-2}\hat{\mathcal{A}}_{-1}
-r_{-1}\{\hat{\mathcal{A}}_{-2}, \hat{\mathcal{A}}_{-2}\}  ) 
\nonumber \\ 
& 
= 2i\bar{\xi} \big( (\partial_{\bar{\xi}} \hat{\mathcal{A}}_{-2})_0 
+(\partial_{\xi} \hat{\mathcal{A}}_{-1})_0
+i\{ \hat{C}, \hat{\bar{C}} \} \big) 
\nonumber \\ 
&\;\quad
-i\xi \big( 2(\partial_{\xi} \hat{\mathcal{A}}_{-2})_0 
+i\{ \hat{\bar{C}}, \hat{\bar{C}} \} \big) 
\nonumber \\ 
&\;\quad
-2i\xi\bar{\xi} \big( 
[ \hat{C}, (\partial_{\xi} \hat{\mathcal{A}}_{-2})_0 ]
+[ \hat{\bar{C}},\, (\partial_{\bar{\xi}} \hat{\mathcal{A}}_{-2})_0 
+(\partial_{\xi} \hat{\mathcal{A}}_{-1})_0 ] \big) 
\,, 
\label{4.21}
\\ 
\hat{\mathcal{F}}_{-2-2-2}&=  
3i( J_{-2-2}\hat{\mathcal{A}}_{-2}
-r_{-2}\{\hat{\mathcal{A}}_{-2}, \hat{\mathcal{A}}_{-2}\} ) 
\nonumber \\ 
& = 3i \bar{\xi} \big( 2(\partial_{\xi} \hat{\mathcal{A}}_{-2})_0 
+i\{ \hat{\bar{C}}, \hat{\bar{C}} \} \big) 
-6i\xi\bar{\xi}\, [\hat{\bar{C}}, (\partial_{\xi} \hat{\mathcal{A}}_{-2})_0 ] 
\,. 
\label{4.22}
\end{align}
Equations (\ref{4.16})--(\ref{4.22}) will be utilized to find a horizontality condition 
appropriate for the present formulation, while Eq. (\ref{4.13})--(\ref{4.15}) 
will be used for examining a Yang-Mills action defined on the supersphere. 


\section{\label{sec:level1}A horizontality condition 
and (anti-)BRST transformations}

This section treats a horizontality condition peculiar to the present formulation. 
It will be shown that the horizontality condition  
put forward by us yields   
the (anti-)BRST transformation rules of the relevant fields on $S_{1}^{n}$.

Now, we posit the condition    
\begin{align}
\hat{\mathcal{F}}_{a\beta\gamma}=0 
\quad \;\;
(\beta, \gamma=-1,-2) \,,
\label{5.1}
\end{align}
which will eventually turn out to be the horizontality condition.  
To begin with, we show that Eq. (\ref{5.1}) yields relations among 
some component fields of $\hat{\mathcal{A}}_{A}$.  
The condition $\hat{\mathcal{F}}_{a\,-1 -1}=0$  
requires that in Eq. (\ref{4.16}), the components of the power series 
in $(\xi, \bar{\xi})$ vanish to yield 
\begin{align}
& (\partial_{\bar{\xi}} \hat{\mathcal{A}}_{-1})_0 =-\frac{i}{2} \{ \hat{C}, \hat{C} \} \,,  
\label{5.2} 
\\
& [\hat{C}, (\partial_{\bar{\xi}} \hat{\mathcal{A}}_{-1})_0 ]=0 \,, 
\label{5.3}
\\
& (\partial_{\bar{\xi}} \hat{\mathcal{A}}_{a})_0 =-i\hat{D}_a \hat{C} 
+ir_a (\partial^2 \hat{\mathcal{A}}_{-1})_0
-ir_a [\hat{C}, (\partial_{\xi} \hat{\mathcal{A}}_{-1})_0 ] \,, 
\label{5.4}
\\
& \hat{D}_a (\partial_{\bar{\xi}} \hat{\mathcal{A}}_{-1})_0 
-\{ \hat{C}, (\partial_{\bar{\xi}} \hat{\mathcal{A}}_{a})_0 \} 
\nonumber \\
& +ir_a \{ \hat{C}, (\partial^2 \hat{\mathcal{A}}_{-1})_0 \} 
+r_a [ (\partial_{\bar{\xi}} \hat{\mathcal{A}}_{-1})_0, 
(\partial_{\xi} \hat{\mathcal{A}}_{-1})_{0} ] =0 \,. 
\label{5.5} 
\end{align}
Obviously Eq. (\ref{5.3}) is satisfied by Eq. (\ref{5.2}).   
Equation (\ref{5.5}) can be found from Eqs. (\ref{5.2}) and (\ref{5.4}). 
Equations (\ref{5.2}) and (\ref{5.4}) are independent of each other, 
as may be seen from their characteristics, such as the ghost numbers and 
the transformation behaviors under the O($n+1$) rotations. 
Hence, it follows that among Eqs. (\ref{5.2})--(\ref{5.5}), 
Eqs. (\ref{5.2}) and (\ref{5.4}) are primary, 
while Eqs. (\ref{5.3}) and (\ref{5.5}) are secondary. 
The condition $\hat{\mathcal{F}}_{a\,-2-2}=0$, together with Eq. (\ref{4.18}),  
leads to    
\begin{align}
& (\partial_{\xi} \hat{\mathcal{A}}_{-2})_0 
=-\frac{i}{2} \{ \hat{\bar{C}}, \hat{\bar{C}} \} \,,
\label{5.6}
\\ 
& (\partial_{\xi} \hat{\mathcal{A}}_{a})_0 =-i\hat{D}_a \hat{\bar{C}} 
+ir_a (\partial^2 \hat{\mathcal{A}}_{-2})_0
+ir_a [\hat{\bar{C}}, (\partial_{\bar{\xi}} \hat{\mathcal{A}}_{-2})_0 ] \,,
\label{5.7}
\\
& [\hat{\bar{C}}, (\partial_{\xi} \hat{\mathcal{A}}_{-2})_0 ]=0 \,, 
\label{5.8}
\\ 
& \hat{D}_a (\partial_{\xi} \hat{\mathcal{A}}_{-2})_0 
-\{ \hat{\bar{C}}, (\partial_{\xi} \hat{\mathcal{A}}_{a})_0 \} 
\nonumber \\
& +ir_a \{ \hat{\bar{C}}, (\partial^2 \hat{\mathcal{A}}_{-2})_0 \} 
+r_a [ (\partial_{\bar{\xi}} \hat{\mathcal{A}}_{-2})_0, 
(\partial_{\xi} \hat{\mathcal{A}}_{-2})_0 ] =0 \,. 
\label{5.9}
\end{align}
Evidently Eqs. (\ref{5.6}) and (\ref{5.7}) are independent of each other. 
Equation (\ref{5.8}) is satisfied by Eq. (\ref{5.6}), while  
Eq. (\ref{5.9}) can be found from Eqs. (\ref{5.6}) and (\ref{5.7}). Hence, it follows that 
among Eqs. (\ref{5.6})--(\ref{5.9}), Eqs. (\ref{5.6}) and (\ref{5.7}) are primary, 
while Eqs. (\ref{5.8}) and (\ref{5.9}) are secondary. 
The condition $\hat{\mathcal{F}}_{a\,-1-2}\,(=\hat{\mathcal{F}}_{a\,-2-1})=0$,  
together with Eq. (\ref{4.17}), gives  
\begin{align}
& (\partial_{\bar{\xi}} \hat{\mathcal{A}}_{-2})_0 
+(\partial_{\xi} \hat{\mathcal{A}}_{-1})_0 =-i\{ \hat{C}, \hat{\bar{C}} \}  \,, 
\label{5.10}
\\
& (\partial_{\bar{\xi}} \hat{\mathcal{A}}_{a})_0 = -i\hat{D}_a \hat{C} 
+ir_a (\partial^2 \hat{\mathcal{A}}_{-1})_0
+ir_a [\hat{C}, (\partial_{\bar{\xi}} \hat{\mathcal{A}}_{-2})_0 ]
+ir_a [\hat{\bar{C}}, (\partial_{\bar{\xi}} \hat{\mathcal{A}}_{-1})_0 ]  \,,
\label{5.11}
\\
& (\partial_{\xi} \hat{\mathcal{A}}_{a})_0 = -i\hat{D}_a \hat{\bar{C}} 
+ir_a (\partial^2 \hat{\mathcal{A}}_{-2})_0
-ir_a [\hat{C}, (\partial_{\xi} \hat{\mathcal{A}}_{-2})_0 ]
-ir_a [\hat{\bar{C}}, (\partial_{\xi} \hat{\mathcal{A}}_{-1})_0 ] \,, 
\label{5.12}
\\ 
& \hat{D}_a (\partial_{\bar{\xi}} \hat{\mathcal{A}}_{-2})_0 
+\hat{D}_a (\partial_{\xi} \hat{\mathcal{A}}_{-1})_0
-\{ \hat{C}, (\partial_{\xi} \hat{\mathcal{A}}_{a})_0 \}
-\{ \hat{\bar{C}}, (\partial_{\bar{\xi}} \hat{\mathcal{A}}_{a})_0 \}
\nonumber \\ 
& +ir_a \{ \hat{C}, (\partial^2 \hat{\mathcal{A}}_{-2})_0 \} 
+ir_a \{ \hat{\bar{C}}, (\partial^2 \hat{\mathcal{A}}_{-1})_0 \} 
+r_a [ (\partial_{\bar{\xi}} \hat{\mathcal{A}}_{-1})_0, 
(\partial_{\xi} \hat{\mathcal{A}}_{-2})_0 ] 
\nonumber \\ 
& -r_a [ (\partial_{\xi} \hat{\mathcal{A}}_{-1})_0, 
(\partial_{\bar{\xi}} \hat{\mathcal{A}}_{-2})_0 ] =0 \,. 
\label{5.13}
\end{align}
It is possible to show that Eq. (\ref{5.11}) reduces to Eq. (\ref{5.4}) by using 
Eqs.  (\ref{5.2}) and (\ref{5.10}), while Eq. (\ref{5.12}) reduces to Eq.(\ref{5.7}) 
by using Eqs. (\ref{5.6}) and (\ref{5.10}). 
Equation (\ref{5.13}) can be found using Eqs. (\ref{5.2}), (\ref{5.4}), (\ref{5.6}), 
(\ref{5.7}), and (\ref{5.10}), and thus it is secondary. 
Hence, it follows that among Eqs. (\ref{5.10})--(\ref{5.13}), 
Eq. (\ref{5.10}) is the only primary relation that we need to consider here. 
We thus conclude that the primary relations obtained from Eq. (\ref{5.1}) are 
essentially five: Eqs (\ref{5.2}), (\ref{5.4}), (\ref{5.6}), (\ref{5.7}), and 
(\ref{5.10}), which are recognized to be independent of each other. 
Conversely, these primary relations lead to Eq. (\ref{5.1}), 
as it is clear from the above investigation.   
Therefore Eq. (\ref{5.1}) is equivalent to a set of the five primary relations,  
which relate the component fields of $\hat{\mathcal{A}}_{A}$,  
except $(\partial^2 \hat{\mathcal{A}}_{a})_0$.

It is easy to see that applying Eqs. (\ref{5.2}), (\ref{5.6}), and (\ref{5.10}) 
to Eqs. (\ref{4.19})--(\ref{4.22}) leads to    
\begin{align}
\hat{\mathcal{F}}_{\alpha\beta\gamma}=0 
\quad \;\; 
(\alpha, \beta, \gamma=-1, -2)\,. 
\label{5.14}
\end{align}
Conversely, Eq. (\ref{5.14}) yields  
Eqs. (\ref{5.2}), (\ref{5.6}), and (\ref{5.10}), 
three of the five primary relations, 
as may be seen from Eqs. (\ref{4.19})--(\ref{4.22}). 
Hence, Eq. (\ref{5.14}) is equivalent to the set of 
Eqs. (\ref{5.2}), (\ref{5.6}), and (\ref{5.10}).
This implies that Eq. (\ref{5.1}) involves Eq. (\ref{5.14}); 
as long as the condition (\ref{5.1}) is taken into account, 
Eq. (\ref{5.14}) becomes redundant. 
As we will see later, Eq. (\ref{5.1}) is confirmed to be the horizontality condition   
appropriate for the present formulation.

Contracting Eq. (\ref{5.4}) by $r_a$ and using Eq. (\ref{2.0}), 
we have 
\begin{align}
(\partial^2 \hat{\mathcal{A}}_{-1})_0
=-ir_a (\partial_{\bar{\xi}} \hat{\mathcal{A}}_{a})_0 
+r_a \hat{D}_a \hat{C} 
+[\hat{C}, (\partial_{\xi} \hat{\mathcal{A}}_{-1})_0 ] \,. 
\label{5.15}
\end{align}
Substituting Eq. (\ref{5.15}) into Eq. (\ref{5.4}) immediately gives 
\begin{align}
P_{ab} (\partial_{\bar{\xi}} \hat{\mathcal{A}}_{b})_0 
=-iP_{ab}  \hat{D}_b \hat{C} \,,
\label{5.16}
\end{align}
with the projection operator $P_{ab}$ defined in Eq. (\ref{2.9}). 
Applying the same procedure to Eq. (\ref{5.7}) leads to 
\begin{align}
(\partial^2 \hat{\mathcal{A}}_{-2})_0
=-ir_a (\partial_{\xi} \hat{\mathcal{A}}_{a})_0 
+r_a \hat{D}_a \hat{\bar{C}} 
-[\hat{\bar{C}}, (\partial_{\bar{\xi}} \hat{\mathcal{A}}_{-2})_0 ] \,, 
\label{5.17}
\end{align}
and 
\begin{align}
P_{ab} (\partial_{\xi} \hat{\mathcal{A}}_{b})_0 
=-iP_{ab}  \hat{D}_b \hat{\bar{C}} \,. 
\label{5.18}
\end{align}

Let us now consider the BRST and anti-BRST transformations,  
denoted by $\boldsymbol{\delta}$ and $\bar{\boldsymbol{\delta}}$ respectively. 
Because the BRST and anti-BRST symmetries are internal symmetries, 
the coordinates $(r_a)$ and each of  $\boldsymbol{\delta}$ and 
$\bar{\boldsymbol{\delta}}$ must commute. 
Accordingly, using the transversality condition (\ref{2.6}), we have 
\begin{align}
& r_a \boldsymbol{\delta} \hat{A}_a =\boldsymbol{\delta} (r_a\hat{A}_a)=0 \,, 
\label{5.19} 
\\ 
& r_a \bar{\boldsymbol{\delta}} \hat{A}_a 
=\bar{\boldsymbol{\delta}} (r_a\hat{A}_a)=0 \,. 
\label{5.20}
\end{align}
They imply that the vectors $\boldsymbol{\delta} \hat{A}_a$ and 
$\bar{\boldsymbol{\delta}} \hat{A}_a$ also live on the tangent space 
${\frak T}_P S^{n}_{1}$. 
The transversality conditions (\ref{5.19}) and (\ref{5.20}) are automatically 
satisfied by setting   
\begin{align}
\boldsymbol{\delta}\hat{A}_a
&\equiv iP_{ab}(\partial_{\bar{\xi}} \hat{\mathcal{A}}_{b})_{0} \,, 
\label{5.21} 
\\
\bar{\boldsymbol{\delta}}\hat{A}_a
&\equiv iP_{ab}(\partial_{\xi} \hat{\mathcal{A}}_{b})_{0} \,.  
\label{5.22}
\end{align}
We can read from Eqs. (\ref{5.21}) and (\ref{5.22}) that the ghost numbers assigned to 
$\boldsymbol{\delta}$ and $\bar{\boldsymbol{\delta}}$ are $1$ and $-1$, respectively 
\cite{KO,NO}.  
Substituting Eq. (\ref{5.16}) into Eq. (\ref{5.21}), 
and Eq. (\ref{5.18}) into Eq. (\ref{5.22}), we have 
\begin{align}
\boldsymbol{\delta}\hat{A}_a &= P_{ab}  \hat{D}_b \hat{C} 
= ir_{b} \mathcal{L}_{ba} \hat{C}  \,,  
\label{5.23} 
\\
\bar{\boldsymbol{\delta}}\hat{A}_a &= P_{ab}  \hat{D}_b \hat{\bar{C}} 
= ir_{b} \mathcal{L}_{ba} \hat{\bar{C}} \,.    
\label{5.24}
\end{align}
Equation (\ref{5.23}) is precisely the BRST transformation rule of $\hat{A}_a$ 
given in Eq. (\ref{2.21});  
in Sec. 2, it was defined by replacing the gauge parameter 
$\lambda$ in Eq. (\ref{2.7}) by the field $\hat{C}$. Equation (\ref{5.24}) describes the 
anti-BRST transformation rule of $\hat{A}_a$, which is nothing but the counterpart 
of Eq. (\ref{5.23}) given by replacing $\hat{C}$ in Eq. (\ref{5.23}) by $\hat{\bar{C}}$. 
Having obtained the expected rules (\ref{5.23}) and (\ref{5.24}), 
the component fields $\hat{A}_a$, $\hat{C}$, and $\hat{\bar{C}}$ 
are confirmed to be the Yang-Mills field, the FP ghost field and the FP anti-ghost field, 
respectively. At the same time, Eqs. (\ref{5.21}) and (\ref{5.22}) are justified.

We next define the BRST and anti-BRST transformations of $\hat{C}$ 
and $\hat{\bar{C}}$ by  
\begin{alignat}{2}
\boldsymbol{\delta}\hat{C} 
& \equiv -(\partial_{\bar{\xi}} \hat{\mathcal{A}}_{-1})_{0} \,, 
\label{5.25}
\\ 
\bar{\boldsymbol{\delta}}\hat{C} 
& \equiv -(\partial_{\xi} \hat{\mathcal{A}}_{-1})_{0} \,, 
\label{5.26}
\\
\boldsymbol{\delta}\hat{\bar{C}} 
& \equiv -(\partial_{\bar{\xi}} \hat{\mathcal{A}}_{-2})_{0} \,,
\label{5.27}
\\
\bar{\boldsymbol{\delta}}\hat{\bar{C}} 
& \equiv -(\partial_{\xi} \hat{\mathcal{A}}_{-2})_{0} \,.
\label{5.28}
\end{alignat}
Then, Eqs. (\ref{5.2}), (\ref{5.6}), and (\ref{5.10}) are written as 
\begin{align}
& \boldsymbol{\delta}\hat{C} =\frac{i}{2} \{ \hat{C}, \hat{C} \} \,,
\label{5.29}
\\ 
& \bar{\boldsymbol{\delta}} \hat{\bar{C}} 
=\frac{i}{2} \{ \hat{\bar{C}}, \hat{\bar{C}} \} \,,
\label{5.30}
\\
& \boldsymbol{\delta}\hat{\bar{C}} +\bar{\boldsymbol{\delta}}\hat{C} 
=i \{ \hat{C}, \hat{\bar{C}} \} \,,
\label{5.31}
\end{align}
respectively. 
They are understood as the (anti-)BRST transformation rules of  
$\hat{C}$ and $\hat{\bar{C}}$. In paticular, Eq. (\ref{5.29}) is identical to  
the BRST transformation rule (\ref{2.22}). 
With Eqs. (\ref{5.23}), (\ref{5.24}), and (\ref{5.29})--(\ref{5.31}), 
it is readily seen that the nilpotency properties  
\begin{subequations}
\label{5.32}
\begin{align}
\boldsymbol{\delta}^2 =\bar{\boldsymbol{\delta}}{}^2=0 
\label{5.32a}
\end{align}
and the anticommutativity property 
\begin{align}
\boldsymbol{\delta}\bar{\boldsymbol{\delta}}
+\bar{\boldsymbol{\delta}}\boldsymbol{\delta}=0 
\label{5.32b}
\end{align}
\end{subequations}
are valid for $\hat{A}_a$, $\hat{C}$, and $\hat{\bar{C}}$. 
In particular, these properties are verified for $\hat{A}_a$ by using 
the property of projection operator $P_{ac} P_{cb}=P_{ab}$. 
In this way, Eqs. (\ref{5.29})--(\ref{5.31}) are confirmed to be the correct 
transformation rules.

What needs to be stressed here that the (anti-)BRST transformation rules 
(\ref{5.23}), (\ref{5.24}), (\ref{5.29}), (\ref{5.30}), and (\ref{5.31}) have 
been derived on the basis of Eq. (\ref{5.1}),  
via the primary relations (\ref{5.4}), (\ref{5.7}), (\ref{5.2}), (\ref{5.6}),  
and (\ref{5.10}), respectively.  
In the ordinary superfield formulation based on the flat superspace 
\cite{sff},  the (anti-)BRST transformation rules of relevant fields are 
derived from the so-called horizontality condition imposed on 
the field strength of the Yang-Mills superfield on the flat superspace. 
Because Eq. (\ref{5.1}) just corresponds to such a condition, we should consider 
Eq. (\ref{5.1}) as the horizontality condition in the present formulation. 
Now we can say that the horizontality condition (\ref{5.1}) yields  
the correct (anti-)BRST transformation rules \footnotemark[6]. 
%
\footnotetext[6]{In the ordinary superfield formulation \cite{sff}, 
the horizontality condition is equivalent to 
the (anti-)BRST transformation rules of relevant fields.  
In contrast, the horizontality condition (\ref{5.1}) is not equivalent
to the transformation rules (\ref{5.23}), (\ref{5.24}), and 
(\ref{5.29})--(\ref{5.31}), because the projection operator 
$P_{ab}$ is used in deriving Eqs. (\ref{5.23}) and (\ref{5.24}); 
we cannot find Eq. (\ref{5.1}) only from 
Eqs. (\ref{5.23}), (\ref{5.24}), and (\ref{5.29})--(\ref{5.31}). 
However, in a later publication \cite{BD2}, we will show that 
Eq. (\ref{5.1}) and the horizontality condition in 
the ordinary superfield formulation are related by a stereographic 
mapping from the supersphere $S_{1}^{n|2}$ to an $(n+2)$-dimensional  
superplane through the use of conformal super Killing vectors.} 

Introducing the Nakanishi-Lautrup field $\hat{B}$,  
we can decompose Eq. (\ref{5.31}) into the two transformation rules: 
\begin{align}
\boldsymbol{\delta}\hat{\bar{C}}& =i\hat{B} \,, 
\label{5.33}
\\
\bar{\boldsymbol{\delta}} \hat{C}& = 
-i\hat{B} +i\{ \hat{C}, \hat{\bar{C}} \} \,.  
\label{5.34}
\end{align}
The (anti-)BRST transformation rules of $\hat{B}$ are defined 
in such a manner that the properties (\ref{5.32}) are valid for $\hat{B}$:  
\begin{align}
\boldsymbol{\delta} \hat{B}& =0 \,, 
\label{5.35}
\\
\bar{\boldsymbol{\delta}} \hat{B}& = 
-i[\hat{B}, \hat{\bar{C}}] \,. 
\label{5.36}
\end{align}

Combining Eq. (\ref{4.7}) with Eq. (\ref{5.21}), 
and Eq. (\ref{4.8}) with Eq. (\ref{5.22}), we have  
\begin{align}
(\partial_{\bar{\xi}} \hat{\mathcal{A}}_{a})_0 
& =-i(\boldsymbol{\delta}\hat{A}_a -r_{a}\hat{C}) \,, 
\label{5.37}
\\
(\partial_{\xi} \hat{\mathcal{A}}_{a})_0 
& =-i(\bar{\boldsymbol{\delta}}\hat{A}_a -r_{a}\hat{\bar{C}}) \,. 
\label{5.38}
\end{align}
Because $\boldsymbol{\delta}\hat{A}_a$ is 
the infinitesimal gauge transformation of $\hat{A}_a$ in a sense,  
Eq. (\ref{5.37}) implies that $(\partial_{\bar{\xi}} \hat{\mathcal{A}}_{a})_0$ 
is decomposed into its own gauge-orbit and radial components. 
Similar decomposition of $(\partial_{\xi} \hat{\mathcal{A}}_{a})_0$ @
is seen in Eq. (\ref{5.38}). 
Substituting Eqs. (\ref{5.26}) and (\ref{5.37}) into Eq. (5.15) and 
using Eqs. (\ref{2.0}), (\ref{4.6}), and (\ref{5.29}), we have 
\begin{align}
(\partial^2 \hat{\mathcal{A}}_{-1})_0 =\hat{C} +r_{\mu} \partial_{\mu} \hat{C}
+i\boldsymbol{\delta} \bar{\boldsymbol{\delta}} \hat{C} \,. 
\label{5.39}
\end{align}
Similarly, substituting Eqs. (\ref{5.27}) and (\ref{5.38}) into Eq. (5.17) and 
using Eqs. (\ref{2.0}), (\ref{4.6}), and (\ref{5.30}),  
we have 
\begin{align}
(\partial^2 \hat{\mathcal{A}}_{-2})_0 =\hat{\bar{C}} 
+r_{\mu} \partial_{\mu} \hat{\bar{C}}
+i\boldsymbol{\delta} \bar{\boldsymbol{\delta}} \hat{\bar{C}} \,. 
\label{5.40}
\end{align}
At this stage, 
all the component fields in Eqs. (4.4), 
except $(\partial^2 \hat{\mathcal{A}}_{a})_0$, 
are written in terms of $\hat{A}_a$, $\hat{C}$, $\hat{\bar{C}}$, and $\hat{B}$. 
(Only the normal component $r_a (\partial^2 \hat{\mathcal{A}}_{a})_0$ can be written 
in terms of these fields; see Eq. (\ref{4.9}).) 
In fact, Eqs. (4.4) can be expressed as  
\begin{subequations}
\label{5.41}
\begin{align}
&\hat{\mathcal{A}}_{a}= \hat{A}_{a} 
-i\bar{\xi}(\boldsymbol{\delta}\hat{A}_a -r_{a}\hat{C})
-i\xi(\bar{\boldsymbol{\delta}}\hat{A}_a -r_{a}\hat{\bar{C}})
+\xi\bar{\xi} (\partial^2 \hat{\mathcal{A}}_{a})_0 \,, 
\label{5.41a}
\\ 
&\hat{\mathcal{A}}_{-1}= -i\hat{C} 
-\bar{\xi} \boldsymbol{\delta}\hat{C}
-\xi \bar{\boldsymbol{\delta}} \hat{C}
+\xi\bar{\xi} 
(\hat{C} +r_{\mu} \partial_{\mu} \hat{C}
+i\boldsymbol{\delta} \bar{\boldsymbol{\delta}} \hat{C})  \,, 
\label{5.41b}
\\ 
&\hat{\mathcal{A}}_{-2}= -i\hat{\bar{C}} 
-\bar{\xi} \boldsymbol{\delta}\hat{\bar{C}} 
-\xi \bar{\boldsymbol{\delta}} \hat{\bar{C}}
+\xi\bar{\xi}
(\hat{\bar{C}} +r_{\mu} \partial_{\mu} \hat{\bar{C}}
+i\boldsymbol{\delta} \bar{\boldsymbol{\delta}} \hat{\bar{C}})  \,. 
\label{5.41c}
\end{align}
\end{subequations}
Note here that the (anti-)BRST transformation rules of 
$\hat{A}_{a}$, $\hat{C}$, $\hat{\bar{C}}$, and $\hat{B}$ are expressed 
in terms of these fields.

In this section, we have treated the tensor components @
$\hat{\mathcal{F}}_{a\beta\gamma}$ 
and $\hat{\mathcal{F}}_{\alpha\beta\gamma}$ to find the 
(anti-)BRST transformation rules of the relevant fields. 
Once the horizontality condition (\ref{5.1}) is set and Eqs. (\ref{5.2})--(\ref{5.13}) 
are derived, the other tensor components $\hat{\mathcal{F}}_{abc}$ and 
$\hat{\mathcal{F}}_{ab\gamma}$, expressed by Eqs. (\ref{4.13})--(\ref{4.15}),  
are extremely simplified, as we just see in the following.  
First, substituting Eqs. (\ref{5.4}) and (\ref{5.7}) into Eq. (\ref{4.13}) 
and using the formula 
$r_a (\partial_b r_c -\partial_c r_b)+\mbox{c.p.}=0$,  
we can simplify Eq. (\ref{4.13}) as 
\begin{align}
\hat{\mathcal{F}}_{abc}&= \hat{F}_{abc} 
-\bar{\xi} [ \hat{F}_{abc}, \hat{C}]  
-\xi  [ \hat{F}_{abc}, \hat{\bar{C}}]  
\nonumber \\ 
&\;\quad +i\xi\bar{\xi} \bigg( -\frac{1}{r_{n+1}} \hat{F}_{ab} \delta_{c(n+1)} 
+\mathcal{L}_{ab}(\partial^{2} \hat{\mathcal{A}}_{c})_0  
\nonumber \\ 
&\;\quad +r_{a} \{ \hat{D}_{b} \hat{C}, \hat{D}_{c} \hat{\bar{C}} \} 
-r_{a} \{ \hat{D}_{c} \hat{C}, \hat{D}_{b} \hat{\bar{C}} \} 
+\mbox{c.p.} \bigg) \,. 
\label{5.42}
\end{align}
Next, substituting Eqs. (\ref{5.2}), (\ref{5.4}), (\ref{5.7}), (\ref{5.25}), and (\ref{5.26}) 
into Eq. (\ref{4.14}) and using the formula 
$(r_a -\delta_{a\mu} r_{\mu}) \delta_{b(n+1)}=(r_b -\delta_{b\mu} r_{\mu}) \delta_{a(n+1)}$,  
we can simplify Eq. (\ref{4.14}) as 
\begin{align}
\hat{\mathcal{F}}_{ab\,-1}&
= i \xi \big( \mathcal{L}_{ab}(\partial_{\xi} \hat{\mathcal{A}}_{-1})_0 
+i \{ \hat{C}, \mathcal{L}_{ab} \hat{\bar{C}} \} 
\nonumber \\ 
&\;\quad
-\hat{F}_{ab} -ir_{a}(\partial^{2} \hat{\mathcal{A}}_{b})_0
+ir_{b}(\partial^{2} \hat{\mathcal{A}}_{a})_0 \big)
\nonumber \\ 
&\;\quad +i\xi\bar{\xi} \big( 
i\mathcal{L}_{ab} \boldsymbol{\delta}\bar{\boldsymbol{\delta}} \hat{C} 
+[ \boldsymbol{\delta} \hat{C}, \mathcal{L}_{ab} \hat{\bar{C}} ]
-[ \bar{\boldsymbol{\delta}} \hat{C}, \mathcal{L}_{ab} \hat{C} ] 
\nonumber \\ 
&\;\quad +[ \hat{F}_{ab} +ir_{a}(\partial^{2} \hat{\mathcal{A}}_{b})_0
-ir_{b}(\partial^{2} \hat{\mathcal{A}}_{a})_0 , \hat{C} ] \big) \,. 
\label{5.43} 
\end{align}
In a similar way,  Eq. (\ref{4.15}) is simplified as  
\begin{align}
\hat{\mathcal{F}}_{ab\,-2}&
= i \bar{\xi} \big( \mathcal{L}_{ab}(\partial_{\bar{\xi}} \hat{\mathcal{A}}_{-2})_0 
+i \{ \hat{\bar{C}}, \mathcal{L}_{ab} \hat{C} \} 
\nonumber \\ 
&\;\quad
+\hat{F}_{ab} +ir_{a}(\partial^{2} \hat{\mathcal{A}}_{b})_0
-ir_{b}(\partial^{2} \hat{\mathcal{A}}_{a})_0 \big)
\nonumber \\ 
&\;\quad +i\xi\bar{\xi} \big( 
i\mathcal{L}_{ab} \boldsymbol{\delta}\bar{\boldsymbol{\delta}} \hat{\bar{C}} 
+[ \boldsymbol{\delta} \hat{\bar{C}}, \mathcal{L}_{ab} \hat{\bar{C}} ]
-[ \bar{\boldsymbol{\delta}} \hat{\bar{C}}, \mathcal{L}_{ab} \hat{C} ] 
\nonumber \\ 
&\;\quad +[ \hat{F}_{ab} +ir_{a}(\partial^{2} \hat{\mathcal{A}}_{b})_0
-ir_{b}(\partial^{2} \hat{\mathcal{A}}_{a})_0 , \hat{\bar{C}} ] \big) \,. 
\label{5.44} 
\end{align}
Here, it should be noted that the zeroth-order terms in $(\bar{\xi}, \xi)$ included 
in Eqs. (\ref{4.14}) and (\ref{4.15}) vanish by using  Eqs. (\ref{5.4}) and (\ref{5.7}),  
and consequently these terms do not appear in Eqs. (\ref{5.43}) and (\ref{5.44}). 
The tensor component $\hat{\mathcal{F}}_{ab\,-1}$ turns out to be proportional to 
$\xi$, while  $\hat{\mathcal{F}}_{ab\,-2}$ turns out to be proportional to 
$\bar{\xi}$.


\section{\label{sec:level1}Yang-Mills actions on supersphere}

In this section, we present a (modified) Yang-Mills action on $S_{1}^{n|2}$  
that eventually turns out to be the Yang-Mills action (\ref{2.15}).

The Yang-Mills action for $\hat{\mathcal{A}}_{A}$ that we think of first is 
a supersymmetric analogue of the action (\ref{2.15}), which is written 
in terms of the field strength $\hat{\mathcal{F}}_{ABC}$ as 
\begin{align}
\mathcal{S}_{\rm YM}
=\int d^{n|2} \varOmega \bigg[ 
{1\over12} \mathrm{Tr}(\hat{\mathcal{F}}^{ABC}
\hat{\mathcal{F}}_{CBA}) 
\bigg] \,, 
\label{6.1}
\end{align}
where $d^{n|2}\varOmega$ is an invariant measure on $S^{n|2}_{1}$ which    
is defined as an analogue of the measure $d^n \varOmega$ given in Eq. (\ref{2.16}): 
\begin{subequations}
\label{6.2}
\begin{align}
d^{n|2} \varOmega 
& \equiv \frac{-i}{\sqrt{(\rho^{n+1})^2}} 
\prod^{n}_{\substack{M=-2 \\ M\neq0}}  d\rho^{M}  
\label{6.2a}
\\
& =\bigg( 1+\frac{i\xi\bar{\xi}}{r_{n+1}^2} \bigg)
d^{n}\varOmega\, id\xi d\bar{\xi} \,. 
\label{6.2b}
\end{align} 
\end{subequations}
Here the expression (\ref{6.2b}) has been obtained by using Eq. (\ref{3.4}).  
The contravariant tensor $\hat{\mathcal{F}}^{ABC}$ is defined by   
\begin{align}
\hat{\mathcal{F}}^{ABC}\equiv (-1)^{|D|(|B|+|E|+|C|+|F|)+|E|(|C|+|F|)}
g^{AD} g^{BE} g^{CF} \hat{\mathcal{F}}_{DEF} \,, 
\label{6.3}
\end{align}
where $g^{AB}$ is the inverse of the metric tensor $g_{AB}$ defined in Eq. (\ref{3.2});  
as a matrix, $g^{AB}$ is the same as $g_{AB}$, i.e., 
$g^{ab}=\delta^{ab}$,~$g^{-1-2}=-g^{-2-1}=-i$.  
The Yang-Mills action $\mathcal{S}_{\rm YM}$ is invariant under 
the transformations specified by the elements of ${\rm OSp}(n+1|2)$ 
(or simply under the ${\rm OSp}(n+1|2)$ transformations).

Now we impose the horizontality condition (\ref{5.1}) on the action $\mathcal{S}_{\rm YM}$. 
Then, the complete ${\rm OSp}(n+1|2)$ symmetry of $\mathcal{S}_{\rm YM}$ is spoiled, 
but only the symmetry specified by the subgroup ${\rm O}(n+1) \times {\rm Sp}(2)$ still remains 
without being spoiled. As was shown in Sec. 5, the condition (\ref{5.1}) involves Eq. (\ref{5.14}). 
For this reason, by setting Eq. (\ref{5.1}),  
the $(\hat{\mathcal{F}})^{2}$-term in Eq. (\ref{6.1}) becomes 
\begin{align}
\hat{\mathcal{F}}^{ABC} \hat{\mathcal{F}}_{CBA}=
-\hat{\mathcal{F}}_{abc} \hat{\mathcal{F}}_{abc}
-6i \hat{\mathcal{F}}_{ab\,-1} \hat{\mathcal{F}}_{ab\,-2} \,. 
\label{6.4}
\end{align}
Furthermore, after substituting Eqs. (\ref{5.42})--(\ref{5.44}) into Eq. (\ref{6.4}), 
the $(\hat{\mathcal{F}})^{2}$-term takes the form 
\begin{align}
\hat{\mathcal{F}}^{ABC} \hat{\mathcal{F}}_{CBA}
= -\hat{F}_{abc} \hat{F}_{abc} 
+\mbox{terms proportional to $\bar{\xi}$ and/or $\xi$} \,.
\label{6.5}
\end{align}
Note here that the $-\hat{F}_{abc} \hat{F}_{abc}$ appears as 
the only zeroth-order term in $(\bar{\xi}, \xi)$, 
because Eqs. (\ref{5.43}) and (\ref{5.44}) 
contain no zeroth-order terms in $(\bar{\xi}, \xi)$. 
As can be seen in the literature on superspace, e.g. Ref. \cite{DeW}, 
the integrations over the real anticommutative numbers $\bar{\xi}$ and 
$\xi$ are defined by  
\begin{align}
\int d\bar{\xi}= \int d\xi =0 \,, \qquad 
\int \bar{\xi}d\bar{\xi} =\int \xi d\xi\ =i \,, 
\label{6.6}
\end{align}
where the fact that $\bar{\xi} d\bar{\xi}$ and $\xi d\xi$ are purely imaginary has been 
taken into account. 
Carrying out the integrations over $\bar{\xi}$ and $\xi$ in Eq. (\ref{6.1}) with Eq. (\ref{6.5}) 
in accordance with Eq. (\ref{6.6}), 
we immediately see that the action $\mathcal{S}_{\rm YM}$ does not reduce to  
the Yang-Mills action $S_{\rm YM}$ given in Eq. (\ref{2.15}).

Inserting $i\bar{\xi} \xi$ into the integrand of Eq. (\ref{6.1}), now 
we modify $\mathcal{S}_{\rm YM}$ as 
\begin{align}
\tilde{\mathcal{S}}_{\rm YM}
=\int d^{n|2} \varOmega\,i \bar{\xi} \xi 
\bigg[ {1\over12} \mathrm{Tr}(\hat{\mathcal{F}}^{ABC}
\hat{\mathcal{F}}_{CBA}) \bigg] \,. 
\label{6.7}
\end{align}
This action is not invariant under the ${\rm OSp}(n+1|2)$ 
transformations any more owing to the insertion of $i\bar{\xi} \xi$;  
it remains invariant only under the ${\rm O}(n+1) \times {\rm Sp}(2)$ transformations.  
After imposing the horizontality condition (\ref{5.1}) on $\tilde{\mathcal{S}}_{\rm YM}$, 
this action reduces to the Yang-Mills action $S_{\rm YM}$ 
by carrying out the integrations over $\bar{\xi}$ and $\xi$.  
Thus, the modified action $\tilde{\mathcal{S}}_{\rm YM}$ is recognized as 
a form of the Yang-Mills action $S_{\rm YM}$.


\section{\label{sec:level1}Gauge-fixing terms}

In this section, we propose two gauge-fixing terms expressed as mass terms for 
the Yang-Mills superfield $\hat{\mathcal{A}}_{A}$. 
One of the two gauge-fixing terms is invariant under the ${\rm OSp}(n+1|2)$ transformations,  
while the other is invariant only under the ${\rm O}(n+1) \times {\rm Sp}(2)$ transformations. 
It will be demonstrated that the ${\rm O}(n+1) \times {\rm Sp}(2)$ invariant gauge-fixing term 
turns out to be a generalization of the gauge-fixing term (\ref{2.20}), 
supplemented with a mass term for the fields $\hat{A}_{a}$, $\hat{C}$, and $\hat{\bar{C}}$.

\subsection{\label{sec:level2}An ${\bf OSp}\boldsymbol{(n+1|2)}$ invariant gauge-fixing term}

As a gauge-fixing term, we first take the (naive) mass term of $\hat{\mathcal{A}}_{A}$:  
\begin{subequations}
\label{7.1}
\begin{align}
\mathcal{S}_{\rm GF} 
&=\int d^{n|2} \varOmega \bigg[ -{1\over2} \mathrm{Tr}( g^{AB} 
\hat{\mathcal{A}}_{B}\hat{\mathcal{A}}_{A}) \bigg] 
\label{7.1a}
\\ 
&=\int d^{n|2} \varOmega \bigg[ -{1\over2} \mathrm{Tr}( 
\hat{\mathcal{A}}_{a}\hat{\mathcal{A}}_{a}) \bigg] 
+\int d^{n|2} \varOmega \big[ -i \mathrm{Tr}( 
\hat{\mathcal{A}}_{-1}\hat{\mathcal{A}}_{-2}) \big] . 
\label{7.1b}
\end{align}
\end{subequations}
It is obvious that this is left invariant under the ${\rm OSp}(n+1|2)$ transformations. 
To begin with, we consider the first term on the right-hand side of Eq. (\ref{7.1b}).  
With Eq. (\ref{5.41a}), it is possible to express 
$\mathrm{Tr}( \hat{\mathcal{A}}_{a}\hat{\mathcal{A}}_{a})$ 
in terms of the component fields $\hat{A}_{a}$, $\hat{C}$, $\hat{\bar{C}}$, and 
$(\partial^2 \hat{\mathcal{A}}_{a})_0$. 
Using Eqs. (\ref{2.0}) and (\ref{2.6}), we obtain  
\begin{align}
\mathrm{Tr}( \hat{\mathcal{A}}_{a} \hat{\mathcal{A}}_{a}) 
& =\mathrm{Tr} \Big[ \hat{A}_{a}\hat{A}_{a}
-i\bar{\xi} \boldsymbol{\delta}(\hat{A}_{a}\hat{A}_{a})
-i\xi \bar{\boldsymbol{\delta}}(\hat{A}_{a}\hat{A}_{a}) 
\nonumber 
\\
& \quad\, +2 \xi\bar{\xi} \Big\{ \hat{A}_{a} (\partial^2 \hat{\mathcal{A}}_{a})_0 
-\boldsymbol{\delta} \hat{A}_{a} \bar{\boldsymbol{\delta}} \hat{A}_{a}
-\hat{C}\hat{\bar{C}} \Big\} \Big] . 
\label{7.2}
\end{align}
As was mentioned under Eq. (\ref{5.40}), 
only the field $(\partial^2 \hat{\mathcal{A}}_{a})_0$, 
apart from its normal component $r_{a} (\partial^2 \hat{\mathcal{A}}_{a})_0$, 
has not been written in terms of 
$\hat{A}_a$, $\hat{C}$, $\hat{\bar{C}}$, and $\hat{B}$. 
Utilizing this remarkable fact, 
we now take the following {\em ansatz} for $(\partial^2 \hat{\mathcal{A}}_{a})_0\,$: 
\begin{align}
(\partial^2 \hat{\mathcal{A}}_{a})_0
=ik \hat{A}_{a} - \boldsymbol{\delta} \bar{\boldsymbol{\delta}} \hat{A}_{a} 
+r_{a} \bigg\{
\frac{i}{r_{n+1}} \hat{A}_{n+1} 
+(\partial_{\xi} \hat{\mathcal{A}}_{-1})_{0}
-(\partial_{\bar{\xi}} \hat{\mathcal{A}}_{-2})_{0} \bigg\} , 
\label{7.3}
\end{align}
where $k$ is a factor to be fixed later.  
This ansatz has been put    
in such a manner that $(\partial^2 \hat{\mathcal{A}}_{a})_0$ satisfies  
the condition (\ref{4.9}) and has the ghost number $0$. 
Substituting Eq. (\ref{7.3}) into Eq. (\ref{7.2}) leads to 
\begin{align}
\mathrm{Tr}( \hat{\mathcal{A}}_{a} \hat{\mathcal{A}}_{a}) 
& =\mathrm{Tr} \Big[ \hat{A}_{a}\hat{A}_{a}
-i\bar{\xi} \boldsymbol{\delta}(\hat{A}_{a}\hat{A}_{a})
-i\xi \bar{\boldsymbol{\delta}}(\hat{A}_{a}\hat{A}_{a}) 
\nonumber 
\\
& \quad\, + \xi\bar{\xi} \Big\{ 
-\boldsymbol{\delta} \bar{\boldsymbol{\delta}} (\hat{A}_{a}\hat{A}_{a}) 
-2\hat{C}\hat{\bar{C}} +2ik \hat{A}_{a}\hat{A}_{a}  \Big\} \Big] . 
\label{7.4}
\end{align}
Then, using Eqs. (\ref{6.2b}) and (\ref{6.6}), the first term on the right-hand side 
of Eq. (\ref{7.1b}) is written     
\begin{align}
\mathcal{S}_{\rm GF1} & \equiv \int d^{n|2} \varOmega \bigg[ -{1\over2} \mathrm{Tr}( 
\hat{\mathcal{A}}_{a}\hat{\mathcal{A}}_{a}) \bigg] 
\nonumber 
\\ 
& = 
\int d^{n} \varOmega \, \mathrm{Tr} \bigg[ \frac{i}{2}
\boldsymbol{\delta} \bar{\boldsymbol{\delta}} (\hat{A}_{a}\hat{A}_{a}) 
+i\hat{C}\hat{\bar{C}}
+\bigg( k+\frac{1}{2r_{n+1}^{2}} \bigg) \hat{A}_{a}\hat{A}_{a} \bigg] . 
\label{7.5}
\end{align}

Next, we consider the second term on the right-hand side 
of Eq. (\ref{7.1b}).  
With Eqs. (\ref{5.41b}) and (\ref{5.41c}), it is possible to express 
$\mathrm{Tr}( \hat{\mathcal{A}}_{-1}\hat{\mathcal{A}}_{-2})$ 
in terms of $\hat{C}$ and $\hat{\bar{C}}$ as follows:   
\begin{align}
\mathrm{Tr}( \hat{\mathcal{A}}_{-1} \hat{\mathcal{A}}_{-2}) 
& =\mathrm{Tr} \Big[ -\hat{C}\hat{\bar{C}} 
+i\bar{\xi} \boldsymbol{\delta} (\hat{C} \hat{\bar{C}})
+i\xi \bar{\boldsymbol{\delta}} (\hat{C} \hat{\bar{C}}) 
\nonumber 
\\
& \quad\, + \xi\bar{\xi} \Big\{ 
\boldsymbol{\delta} \bar{\boldsymbol{\delta}} (\hat{C} \hat{\bar{C}}) 
-2i \hat{C} \hat{\bar{C}} -ir_{\mu} \partial_{\mu} (\hat{C} \hat{\bar{C}}) 
\Big\} \Big] .
\label{7.6}
\end{align}
Then, using Eqs. (\ref{6.2b}) and (\ref{6.6}), the second term on the right-hand side 
of Eq. (\ref{7.1b}) can be written   
\begin{align}
\mathcal{S}_{\rm GF2} & \equiv\int d^{n|2} \varOmega \big[ -i \mathrm{Tr}( 
\hat{\mathcal{A}}_{-1}\hat{\mathcal{A}}_{-2}) \big] 
\nonumber 
\\ 
&= 
\int d^{n} \varOmega \,\mathrm{Tr} \bigg[ 
\boldsymbol{\delta} \bar{\boldsymbol{\delta}} (\hat{C} \hat{\bar{C}}) 
-i\bigg( 2+\frac{1}{r_{n+1}^{2}} \bigg) \hat{C} \hat{\bar{C}} 
-ir_{\mu} \partial_{\mu} (\hat{C} \hat{\bar{C}}) \bigg] 
\nonumber 
\\ 
&= 
\int d^{n} \varOmega \,\mathrm{Tr} \Big[ 
\boldsymbol{\delta} \bar{\boldsymbol{\delta}} (\hat{C} \hat{\bar{C}}) 
+i(n-3) \hat{C} \hat{\bar{C}}  \Big] .
\label{7.7}
\end{align}
Here, an integration by parts has been carried out to obtain the final form. 
The gauge-fixing term $\mathcal{S}_{\rm GF}$ is given as the sum of 
Eqs. (\ref{7.5}) and (\ref{7.7}): 
\begin{align}
\mathcal{S}_{\rm GF}&= \mathcal{S}_{\rm GF1} +\mathcal{S}_{\rm GF2}  
\nonumber
\\
&=
\int d^{n} \varOmega \, \mathrm{Tr} \bigg[ \frac{i}{2}
\boldsymbol{\delta} \bar{\boldsymbol{\delta}} (\hat{A}_{a}\hat{A}_{a}
-2i\hat{C}\hat{\bar{C}} ) 
\nonumber 
\\
&\quad\,
+\bigg( k+\frac{1}{2r_{n+1}^{2}} \bigg) \hat{A}_{a}\hat{A}_{a} 
+i(n-2) \hat{C}\hat{\bar{C}} \bigg] . 
\label{7.8}
\end{align}
In order that  $\mathcal{S}_{\rm GF}$ can be invariant under the 
BRST and anti-BRST transformations 
by virtue of Eqs. (\ref{5.32}), we need to choose $k$ and $n$ as  
\begin{align}
k=-\frac{1}{2r_{n+1}^{2}} \,, \quad n=2 \,.
\label{7.9}
\end{align}
The second one, $n=2$, implies that the procedure that we have followed can be applied 
only to the 2-dimensional case, that is, to the sphere $S^{2}$. 
Also, in Eq. (\ref{7.8}), we do not have the room choosing an arbitrary gauge,  
because the ${\rm OSp}(n+1|2)$ invariance of $\mathcal{S}_{\rm GF}$ forbids 
that $\mathcal{S}_{\rm GF}$ contains gauge parameters. 
This consequence would lead to interesting results, 
but we next consider another possibility to proceed in any dimension 
and to introduce a gauge parameter.

\subsection{\label{sec:level2}An  
$\boldsymbol{{\rm O}(n+1) \!\times\! {\rm Sp(2)}}$ invariant gauge-fixing term}

Now, instead of $\mathcal{S}_{\rm GF}$, we adopt a generalization of  
$\mathcal{S}_{\rm GF}$,  
i.e. a {\em generalized} mass term for $\hat{\mathcal{A}}_{a}$, defined by 
\begin{align}
\tilde{\mathcal{S}}_{\rm GF}
=\tilde{\mathcal{S}}_{\rm GF1}+\tilde{\mathcal{S}}_{\rm GF2} \,, 
\label{7.10}
\end{align}
with
\begin{align}
\tilde{\mathcal{S}}_{\rm GF1}
&\equiv \int d^{n|2} \varOmega \bigg[ 
-{1\over2} (1+i \beta \bar{\xi} \xi) 
\mathrm{Tr}( \hat{\mathcal{A}}_{a}\hat{\mathcal{A}}_{a}) \bigg] ,
\label{7.11}
\\
\tilde{\mathcal{S}}_{\rm GF2}
&\equiv \int d^{n|2} \varOmega \bigg[ 
-\frac{i}{2} (\alpha+ i\gamma \bar{\xi} \xi) \mathrm{Tr}( 
\hat{\mathcal{A}}_{-1}\hat{\mathcal{A}}_{-2}) \bigg] . 
\label{7.12}
\end{align}
Here, $\alpha$, $\beta$ and $\gamma$ are constant parameters;  
later some constraints are imposed among them.  
Owing to the presence of the parameters, the gauge-fixing term 
$\tilde{\mathcal{S}}_{\rm GF}$ is not invariant under the ${\rm OSp}(n+1|2)$ 
transformations and it remains invariant only  
under the ${\rm O}(n+1) \times {\rm Sp}(2)$ transformations. 
If the parameters take the values $\alpha=2$ and $\beta=\gamma=0$, 
then $\tilde{\mathcal{S}}_{\rm GF}$ reduces to $\mathcal{S}_{\rm GF}$, 
so that the ${\rm OSp}(n+1|2)$ invariance is restored. 
(The modification from $\mathcal{S}_{\rm GF}$ to $\tilde{\mathcal{S}}_{\rm GF}$ 
may be understood on the basis of a squashing of $S_{1}^{n|2}$.)  
Substituting Eqs. (\ref{7.4}) and (\ref{7.6}) into Eqs. (\ref{7.11}) and (\ref{7.12}),  
respectively, leads to     
\begin{align}
\tilde{\mathcal{S}}_{\rm GF1}& =\int d^{n} \varOmega \, \mathrm{Tr} \bigg[ \frac{i}{2}
\boldsymbol{\delta} \bar{\boldsymbol{\delta}} (\hat{A}_{a}\hat{A}_{a}) 
+i\hat{C}\hat{\bar{C}}
+\bigg( k+\frac{1}{2r_{n+1}^{2}} -\frac{\beta}{2} \bigg) \hat{A}_{a}\hat{A}_{a} \bigg] , 
\label{7.13}
\\
\tilde{\mathcal{S}}_{\rm GF2}& =\int d^{n} \varOmega \,\mathrm{Tr} \bigg[ \frac{1}{2}
\boldsymbol{\delta} \bar{\boldsymbol{\delta}} (\alpha \hat{C} \hat{\bar{C}}) 
+\frac{i}{2} \{(n-3)\alpha+\gamma\} \hat{C} \hat{\bar{C}}  \bigg] . 
\label{7.14}
\end{align}
Hence $\tilde{\mathcal{S}}_{\rm GF}$ can read  
\begin{align}
\tilde{\mathcal{S}}_{\rm GF}& =\int d^{n} \varOmega \, \mathrm{Tr} \bigg[ \frac{i}{2}
\boldsymbol{\delta} \bar{\boldsymbol{\delta}} (\hat{A}_{a}\hat{A}_{a}
-i\alpha \hat{C}\hat{\bar{C}} ) 
+\frac{\kappa}{2} \hat{A}_{a}\hat{A}_{a} +i \omega \hat{C}\hat{\bar{C}} \bigg] , 
\label{7.15}
\end{align}
where 
\begin{subequations}
\label{7.16}
\begin{align}
\kappa &\equiv 2 \bigg(k+\frac{1}{2r_{n+1}^{2}} -\frac{\beta}{2}\bigg) ,  
\label{7.16a}
\\
\omega &\equiv {1\over2} \{ 2+ (n-3)\alpha+\gamma \} .
\label{7.16b}
\end{align}
\end{subequations}
We now decompose $\tilde{\mathcal{S}}_{\rm GF}$ into 
the BRST and anti-BRST double coboundary part, $S_{\rm C}$,  
and the remainder, $S_{\rm M}$, in such a way that  
\begin{align}
\tilde{\mathcal{S}}_{\rm GF}=S_{\rm C} +S_{\rm M} \,, 
\label{7.17}
\end{align}
with 
\begin{align}
S_{\rm C}& \equiv \int d^{n} \varOmega \,   
\boldsymbol{\delta} \bar{\boldsymbol{\delta}} 
\bigg[ \frac{i}{2} \mathrm{Tr} (\hat{A}_{a}\hat{A}_{a}
-i\alpha \hat{C}\hat{\bar{C}} ) \bigg] ,
\label{7.18}
\\
S_{\rm M}& \equiv \int d^{n} \varOmega \,\mathrm{Tr} \bigg[ 
\frac{\kappa}{2} \hat{A}_{a}\hat{A}_{a} +i \omega \hat{C}\hat{\bar{C}} \bigg] . 
\label{7.19}
\end{align}
Note here that $S_{\rm M}$ is precisely a mass term for 
$\hat{A}_{a}$, $\hat{C}$, and $\hat{\bar{C}}$. 
If the parameters $\kappa$ and $\omega$ are chosen to be 
\begin{align}
\kappa=\omega=0 \,, 
\label{7.20}
\end{align}
then $S_{\rm M}$ vanishes, and consequently  
the gauge fixing term $\tilde{\mathcal{S}}_{\rm GF}$ 
becomes invariant under the BRST and anti-BRST transformations 
by virtue of Eqs. (\ref{5.32}). 
(As demonstrated in the next section, 
$S_{\rm M}$ is not invariant 
under the BRST and anti-BRST transformations.) 
Even after having imposed the condition (\ref{7.20}), 
the space dimension $n$ and the constant $\alpha$, which is regarded as 
a gauge parameter, still remain arbitrary. 
Thus, by virtue of the presence of the constant $\gamma$, 
the difficulty lying in the gauge-fixing term $\mathcal{S}_{\rm GF}$ does not 
arise in $\tilde{\mathcal{S}}_{\rm GF}$.

First carrying out the anti-BRST transformation contained in the right-hand side  
of Eq. (\ref{7.18}) and subsequently carrying out the BRST transformation,  
we have 
\begin{subequations}
\label{7.21}
\begin{align}
S_{\rm C} 
&= \int d^{n} \varOmega\,  i \boldsymbol\delta \mathrm{Tr}
\bigg[ (ir_a L_{ab} \hat{\bar{C}}) \hat{A}_b 
-\frac{\alpha}{2} \hat{\bar{C}} 
\bigg(\hat{B}  -\frac{1}{2} \{ \hat{C}, \hat{\bar{C}} \} \bigg) 
\bigg]  
\label{7.21a} 
\\
&= \int d^{n} \varOmega \, \mathrm{Tr} \bigg[ -(ir_a L_{ab} \hat{B}) \hat{A}_b  
+(ir_a L_{ac} \hat{\bar{C}}) r_{b} \mathcal{L}_{bc} \hat{C} 
\nonumber 
\\
& \quad \,
+\frac{\alpha}{2} \bigg( \hat{B}^2 -\hat{B} \{ \hat{C}, \hat{\bar{C}} \} 
+\frac{1}{2} \{ \hat{C}, \hat{\bar{C}} \}{}^{2} 
\bigg) \bigg] . 
\label{7.21b}
\end{align}
\end{subequations}
Equation (\ref{7.21a}) corresponds to Eq. (\ref{2.20}), but is not exactly same except 
in the Landau gauge $\alpha=0$. 
This can be understood from the fact that $S_{\mathrm{GF}}$ with $\alpha\neq0$ 
can never be expressed in 
a BRST and anti-BRST double coboundary form like Eq. (\ref{7.18}), 
although $S_{\mathrm{GF}}$ can be written as an anti-BRST coboundary 
term \footnotemark[7]. 
%
\footnotetext[7]{The gauge-fixing term $S_{\mathrm{GF}}$ can take the following form:  
$$S_{\rm GF}= \int d^{n} \varOmega \,i \bar{\boldsymbol{\delta}} \mathrm{Tr}
\bigg[ \hat{C} \bigg(ir_a L_{ab} \hat{A}_b +\frac{\alpha}{2} \hat{B} \bigg) 
\bigg] . $$} 
%
Integrating by parts over $(r_{\mu})$ and using Eqs. (\ref{2.0}), (\ref{2.4}) and 
(\ref{2.6}), we can rewrite Eq. (\ref{7.21b}) as 
\begin{align}
S_{\rm C}
& =\int d^{n} \varOmega \, \mathrm{Tr} \bigg[ 
\hat{B} \bigg( ir_a L_{ab}\hat{A}_b  
-\frac{\alpha}{2} \{ \hat{C}, \hat{\bar{C}} \} \bigg)
+\frac{\alpha}{2} \hat{B}^2 
\nonumber 
\\ 
& \quad \,
+(ir_a L_{ac} \hat{\bar{C}}) r_{b} \mathcal{L}_{bc} \hat{C} 
+\frac{\alpha}{4} \{ \hat{C}, \hat{\bar{C}} \}{}^{2}  \bigg] . 
\label{7.22}
\end{align}
Using the formulas (\ref{2.26}) and (\ref{2.29}), 
$S_{\rm C}$ can also be written 
\begin{subequations}
\label{7.23}
\begin{align}
S_{\rm C}
& =\int d^{n} \varOmega \, \mathrm{Tr} \bigg[ 
\hat{B} ir_a L_{ab}\hat{A}_b +\frac{\alpha}{2} \hat{B}^2 
-\frac{1}{2} \hat{\bar{C}} 
\big(iL_{ab} \mathcal{L}_{ab} \hat{C} +\alpha [\hat{C}, B] \big)
+\frac{\alpha}{4} \{ \hat{C}, \hat{\bar{C}} \}{}^{2} \bigg] 
\label{7.23a}
\\
& =\int d^{n} \varOmega \, \mathrm{Tr} \bigg[ 
\hat{B} ir_a L_{ab}\hat{A}_b +\frac{\alpha}{2} \hat{B}^2 
+\frac{1}{2}  \hat{C} \big(i\mathcal{L}_{ab} L_{ab} \hat{\bar{C}} 
-\alpha [ \hat{\bar{C}}, B] \big) 
+\frac{\alpha}{4} \{ \hat{C}, \hat{\bar{C}} \}{}^{2} \bigg] . 
\label{7.23b}
\end{align}
\end{subequations}
Equations (\ref{7.21b}), (\ref{7.23a}), and (\ref{7.23b}) 
correspond to Eqs. (\ref{2.27}), (\ref{2.25}), and (\ref{2.28}), respectively.

From the total action 
\begin{align}
\mathcal{S}=\tilde{\mathcal{S}}_{\rm YM}+\tilde{\mathcal{S}}_{\mathrm{GF}} 
=S_{\rm YM}+S_{\rm C} \,, 
\label{7.24}
\end{align}
the Euler-Lagrange equations for $\hat{A}_a$, $\hat{B}$, $\hat{\bar{C}}$,  
and $\hat{C}$ are derived, respectively, as 
\begin{align}
&{i\over2} \mathcal{L}_{ab} \hat{F}_{abc}
=ir_b L_{bc} \hat{B}-\{ ir_b L_{bc} \hat{\bar{C}}, \hat{C} \} \,,
\label{7.25} 
\\ 
& ir_a L_{ab} \hat{A}_b +\alpha {\hat{B}}^{\prime} 
=0 \,, 
\label{7.26} 
\\
& L_{ab} \mathcal{L}_{ab} \hat{C}
-i\alpha [\hat{C}, {\hat{B}}^{\prime\,}] =0 \,, 
\label{7.27} 
\\
& \mathcal{L}_{ab} L_{ab} \hat{\bar{C}}
+i\alpha [\hat{\bar{C}}, {\hat{B}}^{\prime\,}] =0 \,, 
\label{7.28}
\end{align}
where  
\begin{align}
{\hat{B}}^{\prime} \equiv 
\hat{B}-\frac{1}{2} \{ \hat{C}, \hat{\bar{C}} \} \,.
\label{7.29}
\end{align}
Equation (\ref{7.26}) is slightly different from the gauge-fixing condition (\ref{2.32}) 
except in the Landau gauge $\alpha=0$,  
because Eq. (\ref{7.26}) contains ${\hat{B}}^{\prime}$ in place of $\hat{B}$.  
This difference is not essential and causes no trouble; if necessary, 
this can be avoided by choosing  
the following decomposition of Eq. (\ref{5.31}): 
\begin{align}
\boldsymbol{\delta}\hat{\bar{C}} =i\hat{B} +\frac{i}{2} \{ \hat{C}, \hat{\bar{C}} \}\,, 
\quad 
\bar{\boldsymbol{\delta}} \hat{C}& = 
-i\hat{B} +\frac{i}{2} \{ \hat{C}, \hat{\bar{C}} \} \,.  
\label{7.30}
\end{align}
With this decomposition, Eq. (\ref{2.32}) is derived, instead of Eq. (\ref{7.26}),  
through the same procedure as has been taken to derive Eq. (\ref{7.26}).  
Of course, we can use Eq. (\ref{7.26}) as a suitable gauge-fixing condition 
without any difficulties. 
Equations (\ref{7.27}) and (\ref{7.28}) correspond to Eqs. (\ref{2.33}) and (\ref{2.34}), 
respectively. From Eqs. (\ref{7.25}) and (\ref{7.28}), 
an analog of Eq. (\ref{2.35}) is found to be 
\begin{align}
\mathcal{L}_{ab} L_{ab} \hat{B}
+i\alpha \{ [\hat{\bar{C}}, \hat{B}^{\prime\,} ], \hat{C} \}
=\{ L_{ab} \hat{\bar{C}}, \mathcal{L}_{ab} \hat{C} \} \,.
\label{7.31}
\end{align}
In this way, the BRST gauge-fixing procedure reviewed in 
Sec. 2 is completely covered with the present supersphere formulation.


\section{\label{sec:level1}Curci-Ferrari mass term on sphere}

This section focuses on the mass term $S_{\rm M}$, 
which was assumed to vanish with the condition (\ref{7.20}). 
Here, we leave $S_{\rm M}$ without setting the condition (\ref{7.20}),  
and show that $S_{\rm M}$ can be identified with  
the Curci-Ferrari mass term \cite{CF1,CF2} 
by appropriately extending its definition to the sphere $S_{1}^{n}$.

The BRST and anti-BRST transformations of $S_{\rm M}$ 
are calculated to be  
\begin{align}
\boldsymbol{\delta} S_{\rm M}
&=\int d^{n} \varOmega \, \mathrm{Tr} \Big[ -\kappa\hat{C} 
\Big( ir_{a} L_{ab} \hat{A}_{b} -\frac{\omega}{\kappa} \hat{B}^{\prime} 
\Big) \Big] \,,
\label{8.1}
\\
\bar{\boldsymbol{\delta}} S_{\rm M}
&=
\int d^{n} \varOmega \, \mathrm{Tr} \Big[ -\kappa\hat{\bar{C}}  
\Big( ir_{a} L_{ab} \hat{A}_{b} -\frac{\omega}{\kappa} \hat{B}^{\prime} 
\Big) \Big] \,, 
\label{8.2}
\end{align}
where integration by parts over $(r_{\mu})$ has been applied to the 
$\omega$-independent terms,  
and Eqs. (\ref{2.0}), (\ref{2.4}), (\ref{2.6}), and (\ref{2.8}) have been used. 
Obviously, the right-hand sides of Eqs. (\ref{8.1}) and (\ref{8.2}) do not vanish; 
hence, in a naive sense, $S_{\rm M}$ is not invariant under  
the BRST and anti-BRST transformations. 
However, provided that 
\begin{align}
\alpha=-\frac{\omega}{\kappa} \,,
\label{8.3}
\end{align}
it is possible for the right-hand sides of Eqs. (\ref{8.1}) and (\ref{8.2})  
to vanish with the aid of the gauge-fixing condition (\ref{7.26}).  
Because this condition has been derived as the Euler-Lagrange equation for $\hat{B}$, 
we can say that the mass term 
$S_{\rm M}$ with Eq. (\ref{8.3}), i.e., 
\begin{align}
S_{\rm M}^{\prime} \equiv \int d^{n} \varOmega \, \kappa 
\mathrm{Tr} \bigg[ 
\frac{1}{2} \hat{A}_{a}\hat{A}_{a} -i \alpha \hat{C}\hat{\bar{C}} \bigg] , 
\label{8.4}
\end{align}
is invariant {\em on-shell} under the BRST and anti-BRST transformations. 
In other words, we can say that $S_{\rm M}^{\prime}$ 
remains invariant only in the configuration space submanifold 
in which $\hat{A}_{a}$ and $\hat{B}$ satisfy Eq. (\ref{7.26}). 
Equation (\ref{8.4}) shows that $\hat{A}_{a}$ has the mass $\sqrt{-\kappa}$, 
while $\hat{C}$ and $\hat{\bar{C}}$ have the mass $\sqrt{-\alpha\kappa}$.

The on-shell (anti-)BRST invariance of $S_{\rm M}^{\prime}$ 
suggests that $S_{\rm M}^{\prime}$ is invariant {\em off-shell}  
under the so-called {\em on-shell} BRST and anti-BRST transformations.     
They are defined only in the case $\alpha\neq0$ by eliminating $\hat{B}$ 
from Eqs. (\ref{5.33}) and (\ref{5.34}) using  Eq. (\ref{7.26}). 
More precisely, the on-shell BRST transformation reads 
\begin{subequations}
\label{8.5}
\begin{align}
\boldsymbol{\delta}^{\prime} \hat{A}_a &= P_{ab}  \hat{D}_b \hat{C} 
= ir_{b} \mathcal{L}_{ba} \hat{C}  \,, 
\label{8.5a}
\\
\boldsymbol{\delta}^{\prime} \hat{C} &=\frac{i}{2} \{ \hat{C}, \hat{C} \} \,,
\label{8.5b}
\\
\boldsymbol{\delta}^{\prime} \hat{\bar{C}}
&=\frac{1}{\alpha} r_{a}L_{ab}\hat{A}_{b}
+\frac{i}{2} \{ \hat{C}, \hat{\bar{C}} \} \,,
\label{8.5c}
\end{align}
\end{subequations}
while the on-shell anti-BRST transformation reads 
\begin{subequations}
\label{8.6}
\begin{align}
\bar{\boldsymbol{\delta}}^{\prime}\hat{A}_a 
&= P_{ab}  \hat{D}_b \hat{\bar{C}} 
= ir_{b} \mathcal{L}_{ba} \hat{\bar{C}} \,.    
\label{8.6a}
\\
\bar{\boldsymbol{\delta}}^{\prime} \hat{C}
&=-\frac{1}{\alpha} r_{a}L_{ab}\hat{A}_{b}
+\frac{i}{2} \{ \hat{C}, \hat{\bar{C}} \} \,,
\label{8.6b}
\\
\bar{\boldsymbol{\delta}}^{\prime} \hat{\bar{C}} 
&=\frac{i}{2} \{ \hat{\bar{C}}, \hat{\bar{C}} \} \,.
\label{8.6c}
\end{align}
\end{subequations}
It is easy to verify that $S_\mathrm{M}^{\prime}$ remains  
invariant under the on-shell BRST and anti-BRST transformations: 
$\boldsymbol{\delta}^{\prime} S_\mathrm{M}^{\prime}
=\bar{\boldsymbol{\delta}}^{\prime} S_\mathrm{M}^{\prime}=0$. 
With this property, $S_\mathrm{M}^{\prime}$ is recognized as 
the Curci-Ferrari mass term \cite{CF1,CF2} defined on the sphere $S_{1}^{n}$. 
Thus, it is concluded that the gauge-fixing term (\ref{7.10}) involves 
the Curci-Ferrari mass term on $S_{1}^{n}$.

By choosing the parameters in Eqs. (\ref{7.11}) and (\ref{7.12}) to be  
$\alpha=2$ and $\beta=\gamma=0$,  
Eq. (\ref{7.10}) becomes the mass term (\ref{7.1a}). 
At the same time, 
with these parameter values and Eq. (\ref{7.16b}), 
the condition (\ref{8.3}) fixes $\kappa$ to be 
\begin{align}
\kappa=-\frac{\omega}{\alpha}=-\frac{n-2}{2} \,.
\label{8.6.5}
\end{align}
Then the mass of $\hat{A}_{a}$ is determined to be $\sqrt{(n-2)/2}$, 
and the masses of $\hat{C}$ and $\hat{\bar{C}}$ are determined to be $\sqrt{n-2}\,$; 
this result implies that $\hat{A}_{a}$, $\hat{C}$, and $\hat{\bar{C}}$ are massive 
when the dimension $n$ of space is higher than two. 
It therefore follows that 
the mass term (\ref{7.1a}) yields the Curci-Ferrari mass term 
with definite mass values that depend only on the space dimension.  
(If the radius of the $n$-dimensional sphere is taken to be $R$, the mass values 
are proportional to $R^{-1}\sqrt{n-2}$.)

Now we consider the total action with the condition  (\ref{8.3}): 
\begin{align}
\mathcal{S}=\tilde{\mathcal{S}}_{\rm YM}+\tilde{\mathcal{S}}_{\mathrm{GF}} 
=S_{\rm YM}+S_{\rm C}+S_\mathrm{M}^{\prime} \,. 
\label{8.7}
\end{align}
Eliminating $\hat{B}$ in $S_{\mathrm{C}}$ by using  Eq. (\ref{7.26}) leads to  
the total action written only in terms of  
$\hat{A}_{a}$, $\hat{C}$, and $\hat{\bar{C}}$. We express it as  
\begin{align}
\mathcal{S}^{\prime}=S_{\rm YM}+S_\mathrm{C}^{\prime}+S_\mathrm{M}^{\prime} \,. 
\label{8.8}
\end{align}
This action describes a spherical analog of the Curci-Ferrari model 
\cite{CF1,CF2} (see also \cite{Oji, BSNW, Wsc, TW}).   
The action $\mathcal{S}^{\prime}$ remains invariant under 
the on-shell BRST and anti-BRST transformations, 
because each term in the right-hand of Eq. (\ref{8.8}) is left invariant under 
these transformations. 
However, the on-shell BRST and anti-BRST transformations  
$\boldsymbol{\delta}^{\prime}$ and $\bar{\boldsymbol{\delta}}^{\prime}$ 
do not satisfy the nilpotency and anticommutativity properties (\ref{5.32})  
even after using Euler-Lagrange equations obtained from 
$\mathcal{S}^{\prime}$ \footnotemark[8]. 
%
\footnotetext[8]{The same situation occurs in the Curci-Ferrari model on 
Minkowski space. 
In this model, it is shown that the breakdown of the nilpotency of the on-shell 
BRST transformation causes the unitarity violation of the physical S-matrix 
\cite{CF2}.} 
%
Only in the massless case $\kappa=0$, the properties (\ref{5.32}) hold for 
$\boldsymbol{\delta}^{\prime}$ and $\bar{\boldsymbol{\delta}}^{\prime}$ 
at the on-shell level by means of the Euler-Lagrange equations for  
$\hat{\bar{C}}$ and $\hat{C}$.

Let us return to Eq. (\ref{8.7}). 
Owing to the presence of $S_{\mathrm{M}}^{\prime}$, 
the total action $\mathcal{S}$ is not 
invariant under the original BRST and anti-BRST transformations 
$\boldsymbol{\delta}$ and $\bar{\boldsymbol{\delta}}$.  
Fortunately, it is possible to make $\mathcal{S}$ (anti-)BRST invariant by changing  
Eqs. (\ref{5.35}) and (\ref{5.36}) to 
\begin{align}
\boldsymbol{\delta} \hat{B}& =\kappa \hat{C} \,, 
\label{8.9}
\\
\bar{\boldsymbol{\delta}} \hat{B}& = \kappa \hat{\bar{C}}
-i[\hat{B}, \hat{\bar{C}}] \,. 
\label{8.10}
\end{align}
After this modification, we have  
$\boldsymbol{\delta}(S_{\rm C}+S_\mathrm{M}^{\prime})
=\bar{\boldsymbol{\delta}}(S_{\rm C}+S_\mathrm{M}^{\prime})=0$, 
and hence it follows that $\mathcal{S}$ remains invariant 
under the {\em modified} BRST and anti-BRST transformations. 
However, the nilpotency and anticommutativity properties (\ref{5.32}) 
turn out to be lost due to the above modification. 
(The same trouble takes place also in the corresponding model on Minkowski space, 
leading to the unitarity violation of the physical S-matrix in this model \cite{NO, Oji}.) 
The transformation rules (\ref{8.9}) and (\ref{8.10}) cannot be found from 
the horizontality condition (\ref{5.1}), 
as similar rules cannot be found from the ordinary horizontality condition on the flat space. 
For this reason, the modified 
BRST and anti-BRST transformations should be considered to be outside the scope 
of the current study.  
Of course, it will be interesting to see how the condition (\ref{5.1}) is modified 
so that Eqs. (\ref{8.9}) and (\ref{8.10}) can be derived.


\section{\label{sec:level1}Summary and discussion}

We have developed a superfield approach to the BRST formalism for  
the Yang-Mills theory on the $n$-dimensional unit sphere $S_{1}^{n}$.     
In this approach, the $(n+2)$-dimensional unit supersphere $S_{1}^{n|2}$ was 
employed as a suitable superspace (or supermanifold) so that the manifestly 
${\rm O}(n+1)$ covariance of the Yang-Mills theory on $S_{1}^{n}$ 
can be inherited by the supersymmetric generalization. 
Thereby, the rank-3 field strength tensor (\ref{4.10}) 
was defined successfully, and the horizontality condition 
was expressed in a concise form (\ref{5.1}).  
It was demonstrated that this condition yields the (anti-)BRST transformation rules 
of the Yang-Mills and FP (anti-)ghost fields on $S_{1}^{n}$. 
In particular, the BRST transformation rules found by this method    
are identical to those given in Ref. \cite{BD}. 
It should be noted here that unlike the ordinary horizontality condition, 
the condition (\ref{5.1}) 
is not equivalent to a set of the (anti-)BRST transformation rules,  
because the projection operator $P_{ab}$ is used to 
derive these rules from the condition (\ref{5.1}).

By virtue of the horizontality condition (\ref{5.1}), 
the modified action for the Yang-Mills superfield on $S_1^{n|2}$, i.e. Eq. (\ref{6.7}), 
reduced to the action for the Yang-Mills field on $S_1^{n}$.  
Furthermore, the condition (\ref{5.1}) made it possible to arrange 
a gauge-fixing term on $S_1^{n}$ as a mass term on $S_1^{n|2}$. 
In fact, a generalization of the gauge-fixing term proposed in Ref. \cite{BD},  
i.e. $S_\mathrm{C}$,  
was expressed as the generalized mass term (\ref{7.10}) with the condition (\ref{7.20}). 
In addition, as mentioned above, 
the correct BRST transformation rules were found from the condition (\ref{5.1}). 
For these reasons, we can say that the BRST gauge-fixing procedure for the Yang-Mills theory 
on $S_{1}^{n}$ \cite{BD} is completely covered with the present supersphere formulation. 
It is remarkable that the generalized mass term (\ref{7.10}) reduces to the sum of  
the gauge-fixing term $S_\mathrm{C}$ and the mass term $S_\mathrm{M}$,  
as in Eq. (\ref{7.17}). After setting the condition (\ref{8.3}), the mass term $S_\mathrm{M}$ 
turned out to be the Curci-Ferrari mass term on $S_{1}^{n}$. 
For this reason, it is concluded that the supersphere formulation admits  
the Curci-Ferrari model on sphere \footnotemark[9]. 
%
\footnotetext[9]{Recently, an attempt has been made to formulate the Curci-Ferrari model 
in geometrical terms of superspace \cite{TW},  in which 
a curved superspace was employed to treat the Curci-Ferrari mass term. 
This approach appears to have some technical ideas common with our superfield approach, 
because the supersphere $S_{1}^{n|2}$ is a kind of curved superspace. 
However, unlike the approach in Refs. \cite{TW}, 
our supersphere formulation considered the rotational supersymmetry 
characterized by ${\rm OSp}(n+1|2)$ and made it possible to  
describe the horizontrlity condition in a simple form (\ref{5.1}).} 
%
This formulation goes beyond the standard Curci-Ferrari interpretation  
in the sense that a connection of the mass term with the gauge-fixing term 
is considered based on the generalized mass term.  
It was also shown that the ${\rm OSp}(n+1|2)$ invariant mass term (\ref{7.1a}) yields 
the Curci-Ferrari mass term with the definite mass values proportional to 
$\sqrt{n-2}$. 
In this way, the ${\rm OSp}(n+1|2)$ invariance of the mass term fixes the masses of 
the Yang-Mills and FP (anti-)ghost fields on $S_{1}^{n}$.

As stated above, the gauge-fixing term $S_\mathrm{C}$, 
together with the mass term $S_\mathrm{M}$, 
can be written as the generalized mass term (\ref{7.10}).  
This leads us to the notion that we may choose the mass term of the Yang-Mills field 
as a gauge-fixing term in the Yang-Mills theory. 
This notion is also supported by the fact that the mass term of the Yang-Mills field 
is not gauge-invariant and spoils gauge invariance of the Yang-Mills action, 
as gauge-fixing terms spoil it. 
Actually, there have been a few studies corresponding to our notion \cite{FT, BM}, 
in which the equivalence between the mass term of the Yang-Mills field and 
the ordinary gauge-fixing term was proven at the quantum-theoretical level. 
It would be interesting to investigate this equivalence from the aspect of the 
supersphere formulation.

The manifestly ${\rm O}(n+1)$ covariant formulation of the Yang-Mills theory  
on $S_1^n$ can be done in terms of stereographic coordinates on  
the $n$-dimensional hyperplane $\bar{\bf R}^{n}\equiv {\bf R}^{n} \cup \{\infty\}$  
\cite{Ban, BD}. 
In this approach, the Yang-Mills theory on $S_1^n$ is stereographically projected 
onto that on $\bar{\bf R}^{n}$ through the use of conformal Killing vectors. 
The method of  stereographic projection can be applied to  
the Yang-Mills theory on $S_{1}^{n|2}$ by using conformal {\em super} Killing vectors, 
which map the Yang-Mills superfield on $S_{1}^{n|2}$ to that on 
the $(n+2)$-dimensional superplane $\bar{\bf R}^{n|2}$.  
With the aid of conformal super Killing vectors,  
the horizontality condition in the supersphere formulation, Eq. (\ref{5.1}), 
can be related to the one in the ordinary superfield formulation  
by a stereographic mapping from $S_{1}^{n|2}$ to $\bar{\bf R}^{n|2}$. 
Details of this point will be explained in a forthcoming paper \cite{BD2}. 
The supersphere formulation developed by us would be extended to 
the BRST formalism for the Yang-Mills theories on de Sitter and anti-de Sitter spaces 
\cite{Ban2}, because these spaces are connected with $S^4$ by Wick-like rotations. 
The method of  stereographic projection must be useful also in this extension.

\begin{acknowledgements}
One of the authors (S.D.) would like to thank Prof. K. Fujikawa for his encouragements 
and the members of S. N. Bose National Centre for Basic Sciences for their kind hospitality 
during his stay when a part of this work was done. 
\end{acknowledgements}


\begin{thebibliography}{99}
%
\bibitem{Adl}
S. L. Adler, Phys. Rev. D 6 (1972) 3445; \\ 
S. L. Adler, Phys. Rev. D 8 (1973) 2400; \\ 
S. L. Adler, hep-ph/0505177. 
%
\bibitem{DS}
I. T. Drummond and G. M. Shore, Ann. Phys. 117 (1979) 89;  \\
G. M. Shore, Ann. Phys. 128 (1980) 376. 
%
\bibitem{Sho} 
G. M. Shore, Ann. Phys. 117 (1979) 121. 
%
\bibitem{JR} 
R. Jackiw and C. Rebbi, Phys. Rev. D 14 (1976) 517. 
%
\bibitem{Ore} 
F. R. Ore, Jr., Phys. Rev. D 15 (1977) 470;  \\ 
F. R. Ore, Jr., Phys. Rev. D 16 (1977) 2577. 
%
\bibitem{NS} 
N. K. Nielsen and B. Schroer, Nucl. Phys. B 127 (1977) 493; \\ 
P. J. O'Donnell and B. Wong, Phys. Lett. B 138 (1984) 274. 
%
\bibitem{Ban} 
R. Banerjee, Ann. Phys. 311 (2004) 245, 
arXiv:hep-th/0307296. 
%
\bibitem{BD} 
R. Banerjee and S. Deguchi, Phys. Lett. B 632 (2006) 579, 
arXiv:hep-th/0509161. 
%
\bibitem{KO} T. Kugo and I. Ojima, Prog. Theor. Phys. Suppl No.66 (1979) 1. 
%
\bibitem{KU} T. Kugo and S. Uehara, Nucl. Phys. B 197 (1982) 378. 
%
\bibitem{NO} N. Nakanishi and I. Ojima, Covariant Operator Formalism of 
Gauge Theories and Quantum Gravity, World Scientific, Singapore, 1990.  
%
\bibitem{NL} N. Nakanishi, Prog. Theor. Phys. 35 (1966) 1111; \\ 
B. Lautrup, K. Dan. Vidensk. Selsk. Mat. Fys. Medd. 35, (11) (1967). 
%
\bibitem{ND}
S. Naka and S. Deguchi, Prog. Theor. Phys. 76 (1986) 1135; \\ 
S. Deguchi and S. Naka, Prog. Theor. Phys. 79 (1988) 209. 
%
\bibitem{susp}
G. Landi, 	Differ. Geom. Appl. 14 (2001) 95, 
arXiv:math-ph/9907020; 
\\ 
K. Hasebe and Y. Kimura, Nucl. Phys. B 709 (2005) 94, 
arXiv:hep-th/0409230; 
\\
A. F. Schunck, C. Wainwright, J. Math. Phys. 46 (2005) 033511, 
arXiv:hep-th/0409257. 
%
\bibitem{FK}
P. G. O. Freund and I. Kaplansky, J. Math. Phys. 17 (1976) 228; \\  
P. G. O. Freund, J. Math. Phys. 17 (1976) 424; \\
V. Rittenberg and M. Scheunert, J. Math. Phys. 19 (1978) 709. 
%
\bibitem{DeW}
B. DeWitt, Supermanifolds, Cambridge Monographs on Mathematical Physics, 
Cambridge University Press, 1984. 
%
\bibitem{sff}
L. Bonora and M. Tonin, Phys. Lett B 98 (1981) 48; \\ 
L. Bonora, P. Pasti and M. Tonin, Nuovo Cimento A 64, (1981) 307; \\
A. C. Hirshfeld and H. Leschke, Phys. Lett. B 101, (1981) 48; \\
R. Delbourgo and P. D. Jarvis, J. Phys. A 15, (1982) 611; \\
L. Baulieu and J. Thierry-Mieg, Nucl. Phys. B 197, (1982) 477; \\
L. Baulieu, Phys. Rep. 129, (1985) 1; \\
S. Deguchi, Mod. Phys. Lett. A 4 (1989) 2625; \\ 
R. P. Malik, Int. J. Mod. Phys. A 23 (2008) 3685, 
arXiv:0704.0064 [hep-th]; \\
R. P. Malik and B. P. Mandal, arXiv:0709.2277 [hep-th]. 
%
\bibitem{Fuj1}
K. Fujikawa, Prog. Theor. Phys. 59 (1978) 2045; \\
K. Fujikawa, Prog. Theor. Phys. 63 (1980) 1364.  
%
\bibitem{JMD}
S. D. Joglekar, Phys. Rev. D 43 (1991) 1307; 
Erratum-ibid. D 48 (1993) 1878; \\ 
S. D. Joglekar and B. P. Mandal, Z. Phys. C 70 (1996) 673; \\ 
S. Deguchi and B. P. Mandal, Mod. Phys. Lett. A 15 (2000) 965,  
arXiv:hep-th/9905167. 
%
\bibitem{BD2}
R. Banerjee and S. Deguchi, in preparation. 
%
\bibitem{CF1}
G. Curci and R. Ferrari, Nuovo Cim. A 32 (1976) 151. 
%
\bibitem{CF2}
G. Curci and R. Ferrari, Phys. Lett. B 63 (1976) 91; \\ 
G. Curci and R. Ferrari, Nuovo Cim. A 35 (1976) 1;  
Erratum-ibid. A 47 (1978) 555. 
%
\bibitem{Oji}
I. Ojima, Z. Phys. C 13 (1982) 173. 
%
\bibitem{BSNW}
J. de Boer, K. Skenderis, P. van Nieuwenhuizen, and A. Waldron, 
Phys. Lett. B 367 (1996) 175, arXive:hep-th/9510167. 
%
\bibitem{Wsc}
N. Wschebor, Int. J. Mod. Phys. A 23 (2008) 2961,  
arXiv:hep-th/0701127. 
%
\bibitem{TW}
M. Tissier and N. Wschebor, Phys. Rev. D 79 (2009) 065008, 
arXiv:0809.1880 [hep-th]; \\
M. Tissier and N. Wschebor, arXiv:0901.3679 [hep-th]. 
%
\bibitem{FT}
K. Fujikawa and H. Terashima, Nucl. Phys. B 577 (2000) 405,  
arXiv:hep-th/9912253; \\
K. Fujikawa and H. Terashima, Int. J. Mod. Phys. A 16 (2001) 1775, 
arXiv:hep-th/0004190. 
%
\bibitem{BM}
R. Banerjee and B. P. Mandal, Phys. Lett. B 488 (2000),  
arXiv:hep-th/0007092. 
%
\bibitem{Ban2}
R. Banerjee, Ann. Phys. 322 (2007) 2129, arXiv:hep-th/0608045; \\ 
R. Banerjee and B. R. Majhi, Ann. Phys. 323 (2008) 705, arXiv:hep-th/0703207. 


\end{thebibliography}
\end{document}